\documentclass[11pt]{article}
\usepackage{newpasp,epsf,times,mathptm}

\def\msun{{\rm\,M_\odot}}
\def\msun{{\rm\,M_\odot}}
\def\kpc{{\rm\,kpc}}
\def\spose#1{\hbox to 0pt{#1\hss}}
\def\lta{\mathrel{\spose{\lower 3pt\hbox{$\mathchar"218$}}
     \raise 2.0pt\hbox{$\mathchar"13C$}}}
\def\gta{\mathrel{\spose{\lower 3pt\hbox{$\mathchar"218$}}
     \raise 2.0pt\hbox{$\mathchar"13E$}}}

\begin{document}

\title{The Milky Way as a Key to Structural Evolution in Galaxies}

\author{Martin D. Weinberg}
\affil{Department of Physics \& Astronomy, University of
  Massachusetts, Amherst, MA 01003-4525, USA}

\begin{abstract}
  Much of our effort in understanding the long-term evolution and
  morphology of the Milky Way and other galaxies has focused on the
  equilibrium of its luminous disk. However, the interplay between all
  components, seen and unseen, is a major cause of observed features
  and drives evolution. I will review the key underlying dynamics, and
  in a number of examples, show how this leads to lopsidedness and
  offset nuclei, may trigger bars and cause warps.  Indeed, the Milky
  Way like most spiral galaxies show exhibit many of these features.
  In addition, the mechanisms suggest that observed morphology depends
  on the properties of the galaxy and only weakly on any particular
  disturbance.  Because of this convergence, understanding a galaxy's
  history will be subtle and require the level of detail that study of
  the Milky Way can provide.
\end{abstract}

\section{Introduction}

In introductory and graduate astronomy textbooks alike, the Milky Way
is used as an introduction to galaxies in general; clearly Galactic
astronomy and the astronomy of galaxies has evolved together.  Most
recently, the study of evolution and morphology at high redshift, e.g.
motivated by studies of the Hubble Deep Field and the Medium Deep
Survey are leading to revisions of the standard paradigms for galaxy
morphology and evolution.  Ironically, it appears that one legacy of
the HST may be abandonment (at least in part) of the Hubble galaxy
classifications (e.g. Conselice et al. 2000).

At the same time, computational technology is nearly keeping up with
observational technology.  Computation allows complex non-linear
prediction through simulations and statistical inference on large data
sets.  The scientific questions remain familiar:
\begin{itemize}
\item Where does structure in galaxies come from?
\item What excites star formation?
\item Where is the mass in a galaxy?
\item What is its role in causing the observed structure?
\end{itemize}
What is increasingly clear is that any individual galaxy is
morphologically variable.  For example, lopsided ($m=1$) asymmetries
are transient (with gigayear timescales), bars may grow slowly or
suddenly and, under some circumstances may decay as well.  Recent work
shows that stellar populations, star formation rates and color depend
on asymmetry (e.g. Rudnick, Rix \& Kennicut 2000).  Because the
properties of a galaxy depend on its history, an understanding of
galaxy evolution requires that we explore the possible scenarios and
underlying mechanisms.  This same goal motivated the dynamicists of
the past several decades and led to our current understanding of
density waves, instability and stability criteria.  Unfortunately,
gravitational dynamics is also conspiring against our investigation of
long-term evolution as I will describe.  In short, a galaxy responds
to a wide variety of stimuli and over many gigayears the historical
record of specific events fades away.  Because state-of-the-art and
planned surveys will allow us to study the global structure Milky Way
in detail, our Galaxy will be key to this understanding.

The overall plan of this talk is a follows.  I will begin with a
cartoon review of galaxy dynamics, emphasizing asymmetries, since it
is through asymmetries that a galaxy evolves.  I will illustrate this
with a review of asymmetries in the Milky Way.  We will see that the
role of satellites and dwarfs may be significant.  Finally, I will
briefly list the role of upcoming missions and surveys. 

\section{A cartoon review of galactic dynamics}

Fundamentally, all asymmetric structure in stellar systems is
transient.  Although early work in galactic structure emphasized
modes, this same work showed that these modes were different than the
familiar discrete modes of plucked strings and drumheads.  Systems
with very large numbers of degrees of freedom have continuum modes, in
addition the more familiar discrete modes, which allow for both global
redistribution and dissipation.  A physicist might say: ``Spiral arms
and bars are a galaxy's way of attempting to reach its minimum energy
state.''  The dynamicist might say: ``Any feature with non-zero
pattern speed must damp through resonance with the stellar orbits.''
Either way, the end result is transient features which change the
morphology of a galaxy.  Let me give you three familiar examples and
one similar but less familiar case.

\subsection{Swing amplification}

Swing amplification, a widely used dynamical model for understanding
the structure and ubiquity of arms, contains all of these elements.
The following cartoon (based on Toomre 1980 and Binney \& Tremaine
1987) of this mechanism first described by Goldreich \& Lynden-Bell
(1965) illustrates the basic mechanisms.

\begin{figure}[thb]
\centering
\epsfxsize=3.5in\epsfbox{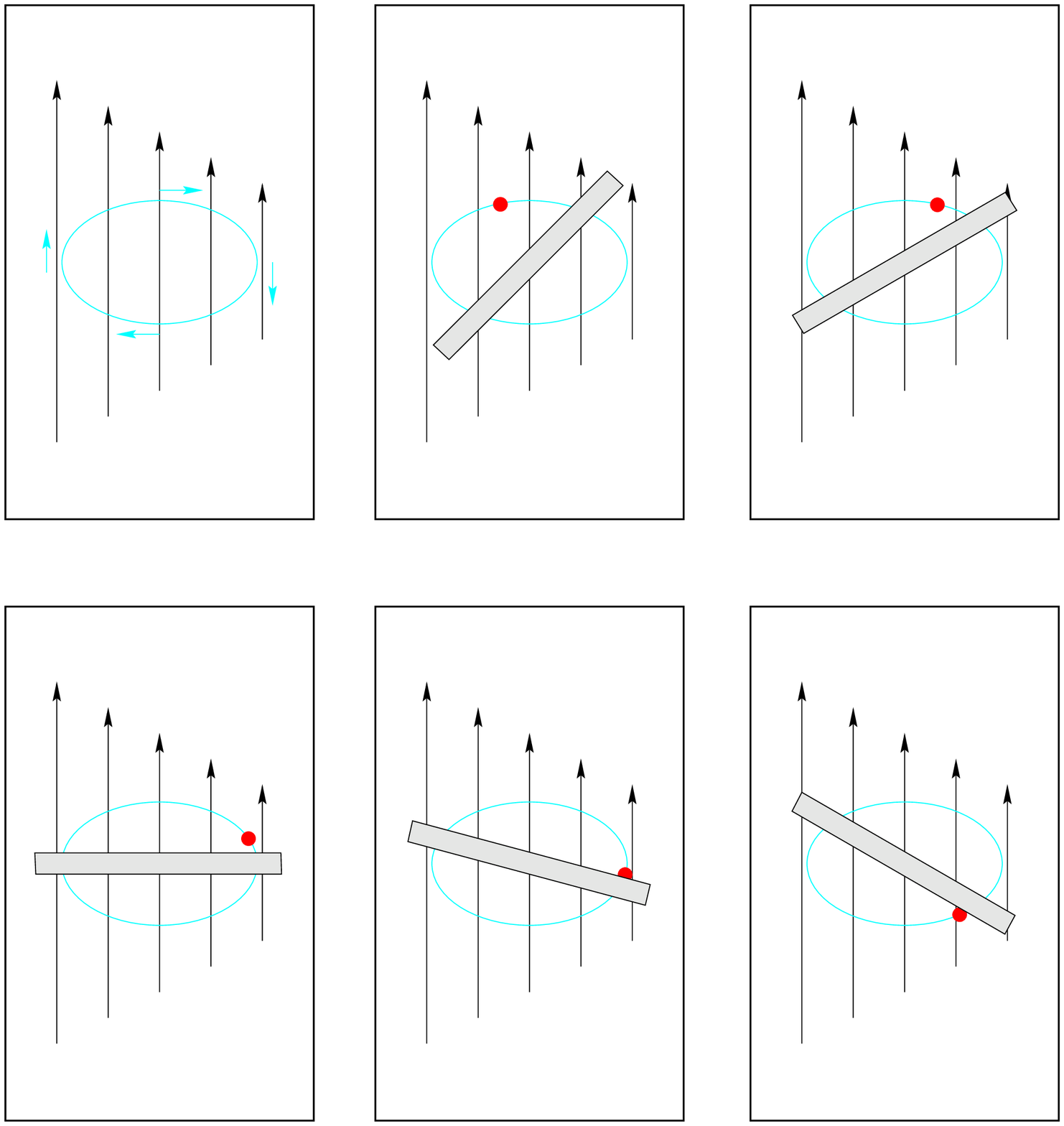}
\caption{Cartoon illustrating swing amplification in a shearing
  sheet (following Binney \& Tremaine 1987).  The vertical arrows
  indicate the relative velocity of the local LSR.  The epicyclic
  orbit in the LSR of the central trajectory is an ellipse and the
  position of the star is the filled dot (first panel).  Successive
  panels show the time evolution of a leading spiral perturbation.
  The trajectory of the star tracks and reinforces the perturbation.}
\label{fig:swing}
\end{figure}

If one takes a small patch out of a galactic disk, it will look like
the situation in the first frame in Figure \ref{fig:swing}.  The LSRs
of orbits appear to lead (trail) as one looks inward (outward) from
the center of the galaxy.  The epicyclic motion about the LSR of an
individual orbit is shown.  Now assume that a leading material arm
appears (2nd frame).  We follow an individual star (filled dot) as
this distribution evolves.  The material arm shears out (successive
frames) but the star feels the excess gravitational attraction of the
arm as it shears, amplifying the arm.

A nice simulation of this has recently been reported by (Demleitner
1998) for both stars and gas together.  The imposed leading distortion
``swings'' quickly and amplitude of the trailing arm increases and
then damps away.
  
Swing amplification is a simple example of transient behavior induced
by a gravitational perturbation.  In the end, a quiet isolated galaxy
may attain a new local axisymmetric state but with a slightly
different equilibrium state since the arm has transferred angular
momentum from some stars to others.  The shape of the resulting arm
depends more on the underlying dynamics of the disk (e.g. Oort
constants) than the details of the triggering distortion.

\subsection{Bar formation}
\label{sec:bar}

Dynamicists argue whether bars form suddenly, as in the sort of
instability that led Ostriker \& Peebles (1972) to advocate a dark
halo, or slowly over many rotation times (Pasha \& Polyachenko 1993).
Either way, the mechanism has a similar explanation.

\begin{figure}
\centering
\mbox{
  \mbox{\epsfxsize=1.0in\epsfbox{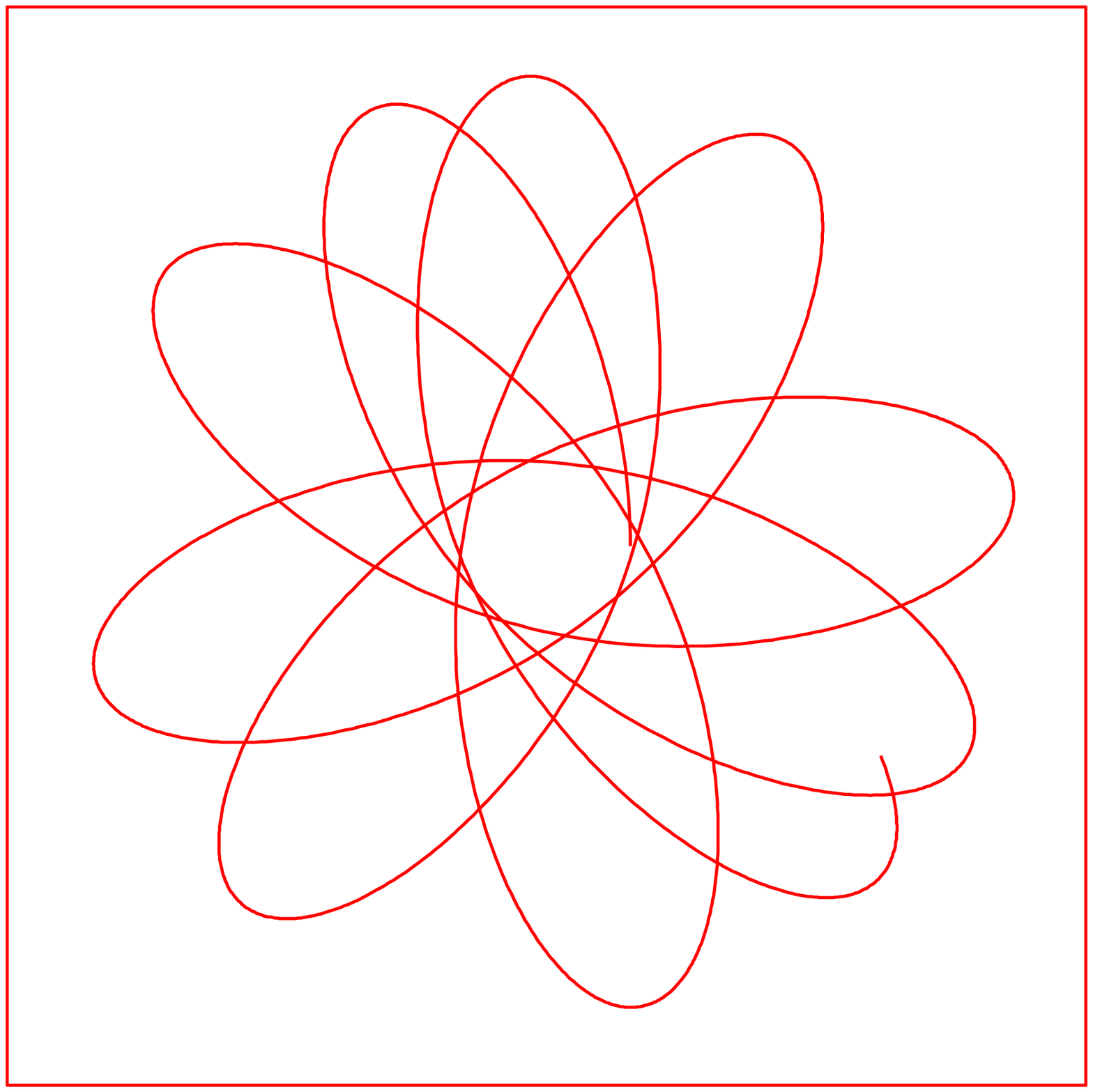}}
  \mbox{\epsfxsize=1.0in\epsfbox{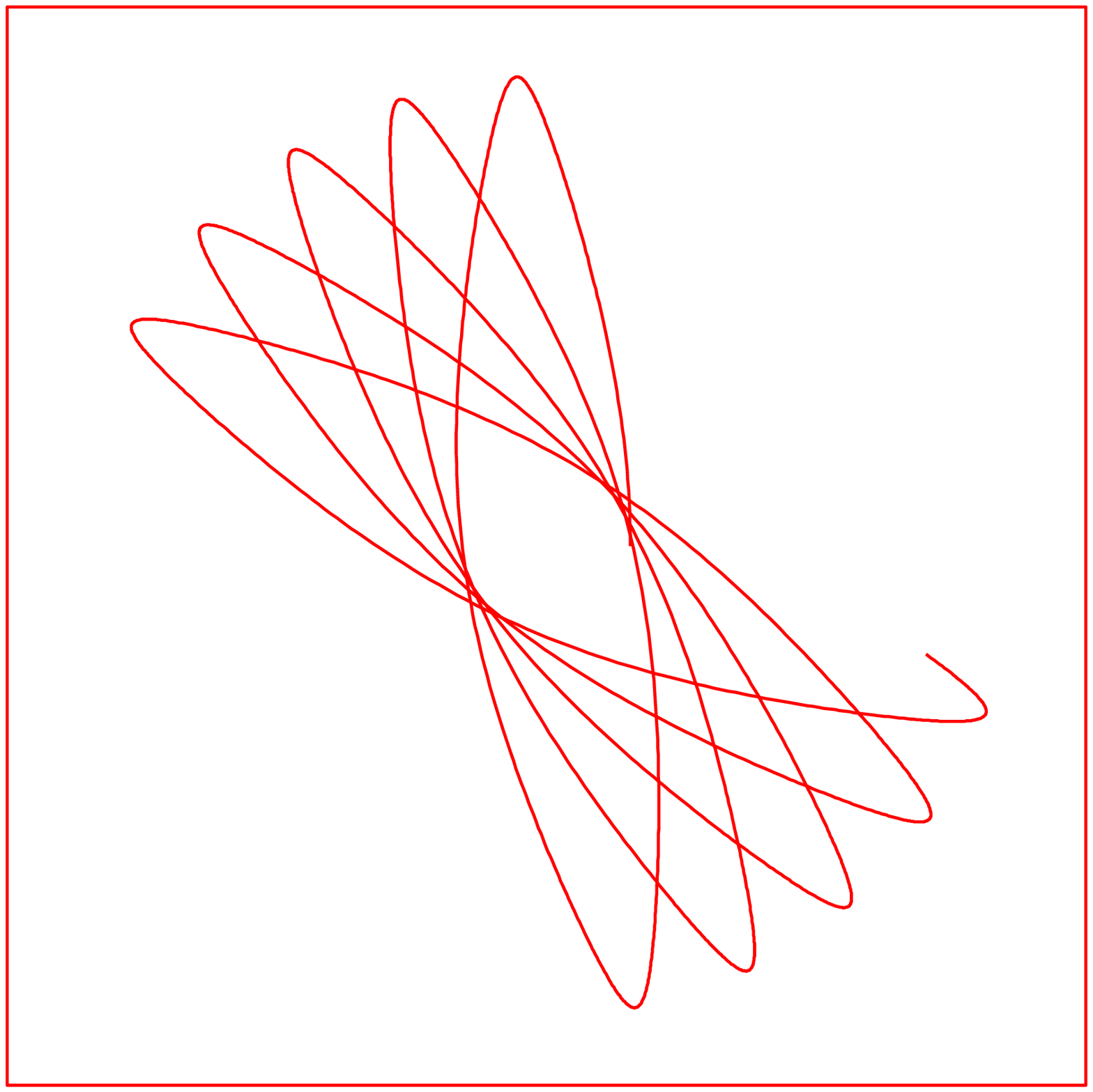}}
  \mbox{\epsfxsize=1.0in\epsfbox{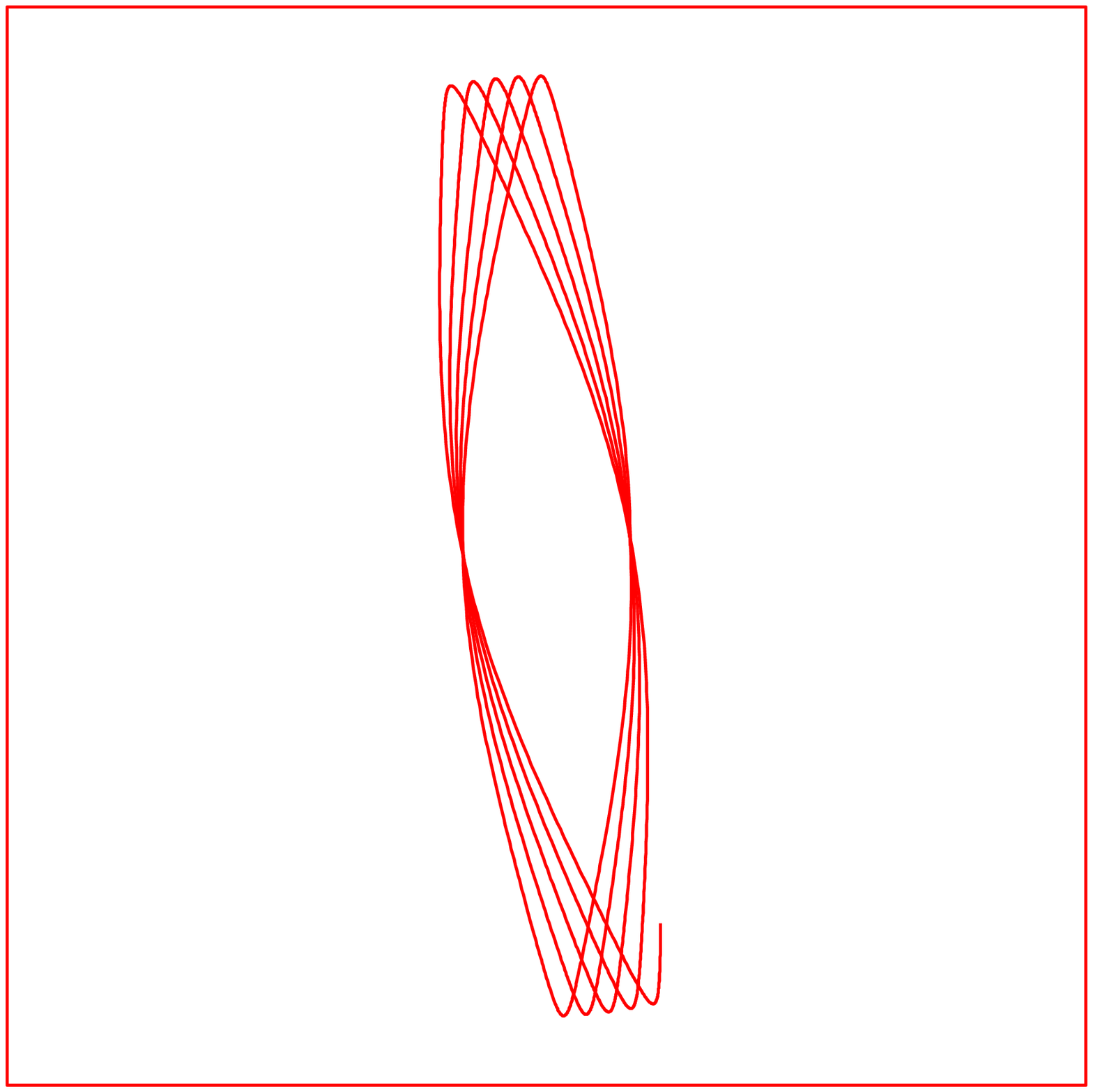}}
  \mbox{\epsfxsize=1.0in\epsfbox{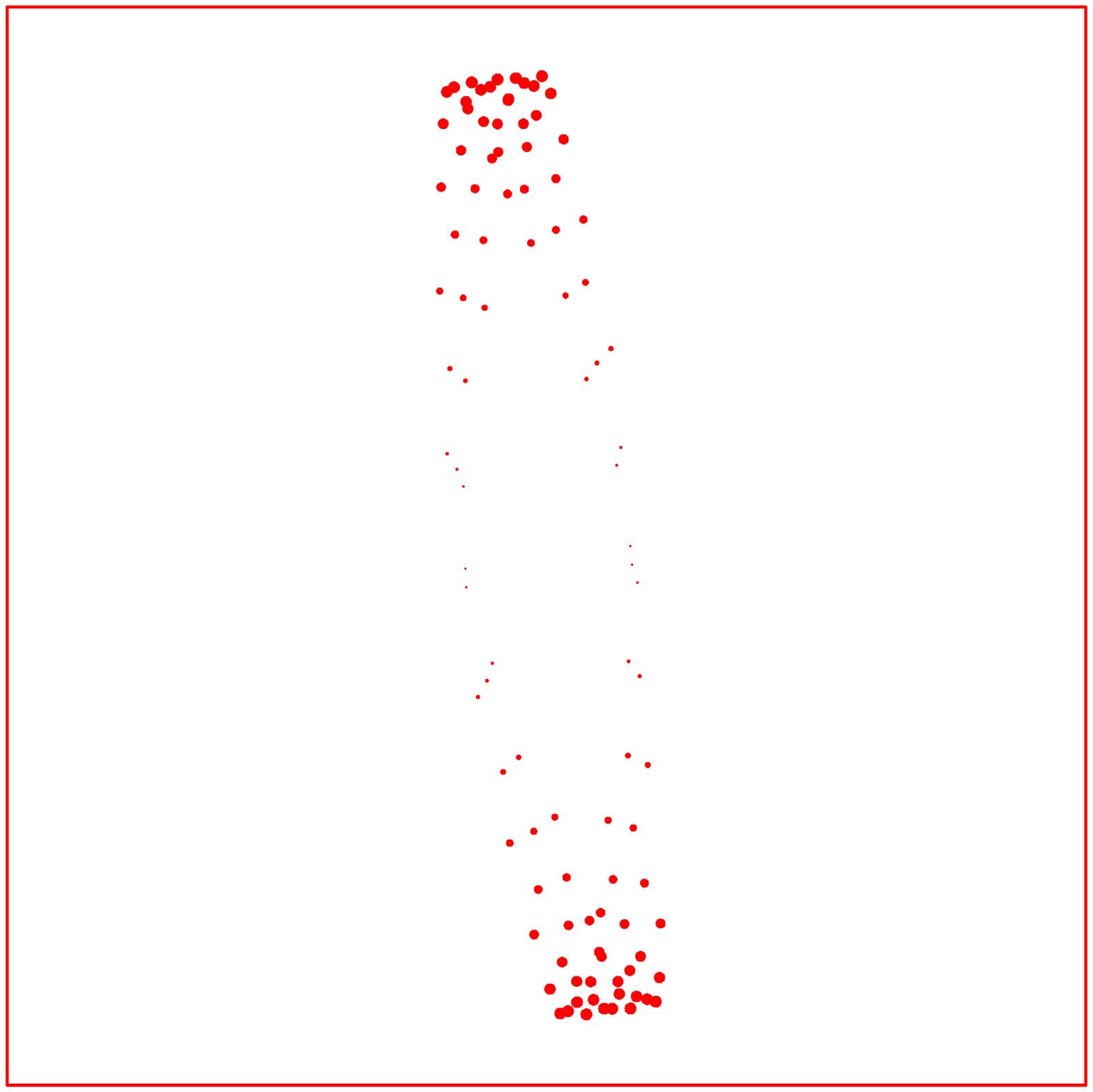}}
  \mbox{\epsfxsize=0.843in\epsfbox{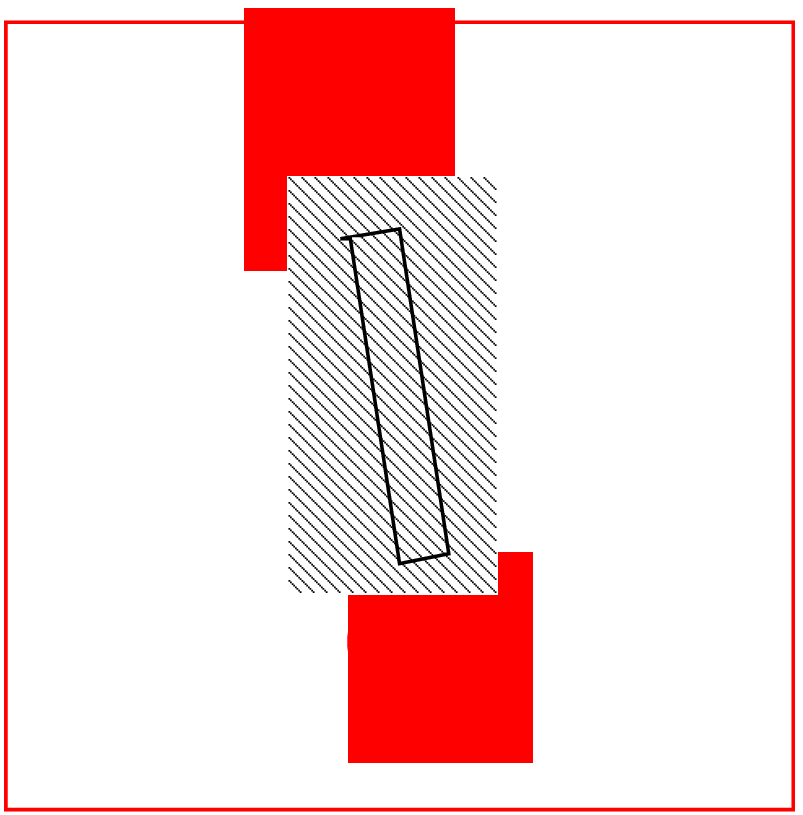}}
  }
\epsfxsize=2.6in\epsfbox{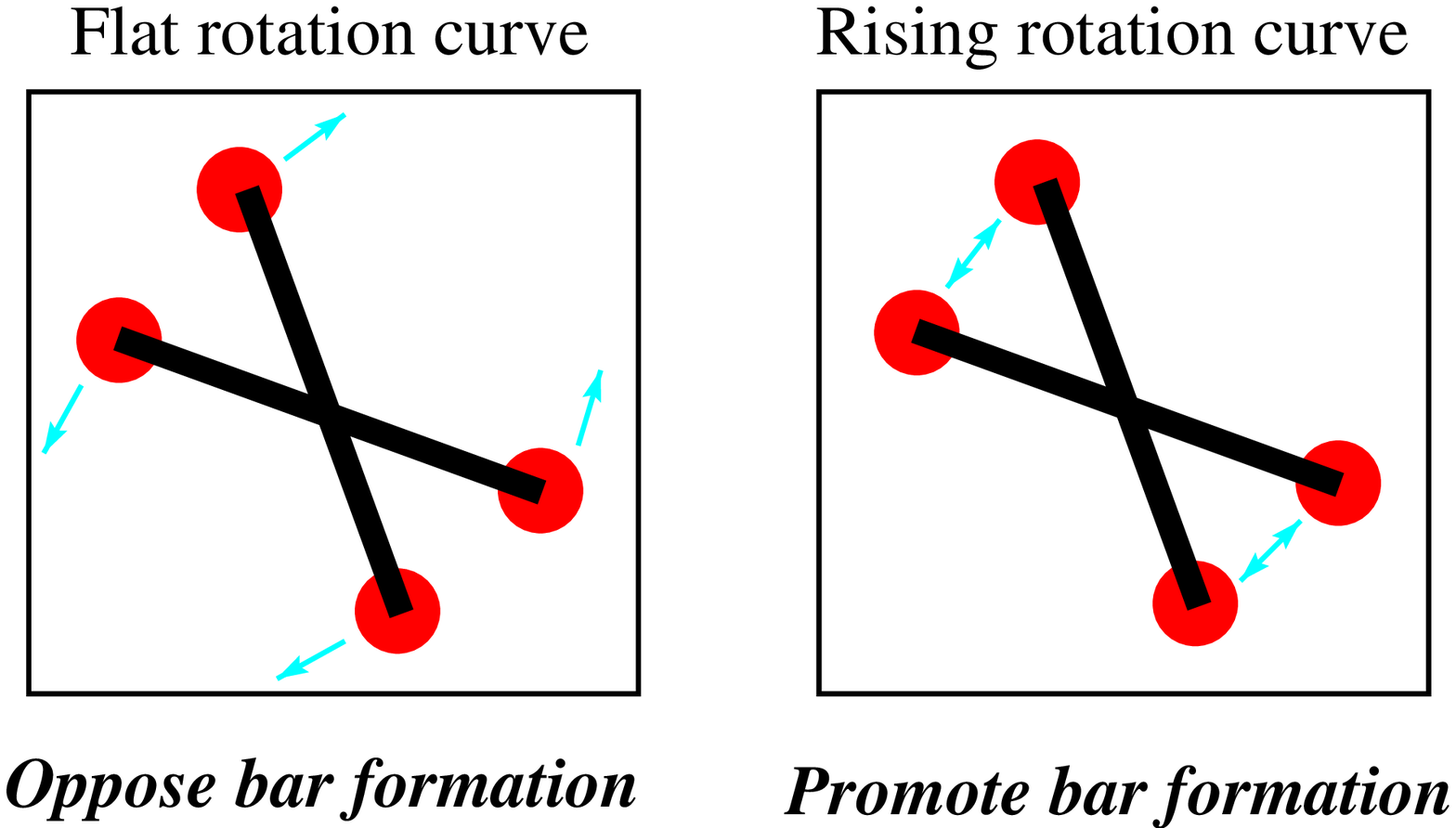}

\caption{Cartoon illustrating bar instability and growth.  Top row:
  orbit approaching resonance in first three panels; fourth panel
  shows the location of a star at fixed time intervals showing that
  the orbit has higher time-averaged space density at apocenter and
  can be represented by a dumbbell (fifth panel).  Bottom row: excess
  density repels apocenters for a flat rotation curve but attracts for
  rising rotation curve, causing bar to grow.}
\label{fig:bar}
\end{figure}

A general orbit in an axisymmetric disk is a rosette.  However, by
accelerating an observer around the disk at a particular angular
velocity, the orbit can be made to nearly close (Fig. \ref{fig:bar}).
Averaged over time, the orbit appears to hang out at its outer turning
points.  You might imagine, therefore, replacing the orbit's average
mass density by a dumbbell (final panel).

Now imagine two such orbits.  If this was a Cavendish experiment, the
ends of the dumbbells would attract.  However, recall that these are {\em
  real} orbits and we are observing them in a rotating frame.
Depending on the background potential, the material torque can cause
the two orbits to precess toward either each other or away from each other.
Lynden-Bell \& Kalnajs (1972) give a precise criteria for this but the
general rule is that orbits precess toward (away from) each if the
rotation curve is rising (flat).

As the process continues, the gravitational potential from the pile of
apoapses increases the influence or {\it reach} of the torque and more
orbits can be influenced to join the crowd.  This is clearly the
description of an instability and it leads to the formation of an
$m=2$ bar.  Thinking like a physicist again, it is favorable for the
bar to exist because the particles that take part lower the energy of
their configuration.  In this way, bar formation may be understood by
analogy with self-gravitating fluid bodies (e.g. Chandrasekhar 1969).
In the presence of a bit of dissipation, a rapidly rotating spheroid
will adopt a prolate shape because it can do so by lowering its
energy.  Similarly, energy (and angular momentum) is redistributed by
the formation of a bar.  Again, the location and structure of the bar
has more to do with the galactic potential and disk distribution than
the perturbation which started off the instability.  Once the bar has
formed, it may continue to influence the evolution of a galaxy; the
existence of a strong non-axisymmetric structure can continue advect
angular momentum both to the disk and halo, acting as an angular
momentum antenna (see Debattista \& Sellwood 1998 for a recent
discussion).  In summary, the existence of a bar implies a process of
continuing evolution.

\subsection{Halo modes and noise}

Now, a gravitational halo itself can sustain modes, and not unlike
those of a disk, these modes damp.  These are less well-known because
they are not practical to observe (however, see the end of this talk)
but can be worked out and detected in n-body simulations.  A halo
dominates the outer galaxy and is significant in the inner galaxy and
therefore persistent halo distortions can be quite important even
though these can not be observed directly.

\begin{figure}
  \centering
  \mbox{
    \mbox{\epsfxsize=2.4in\epsfbox{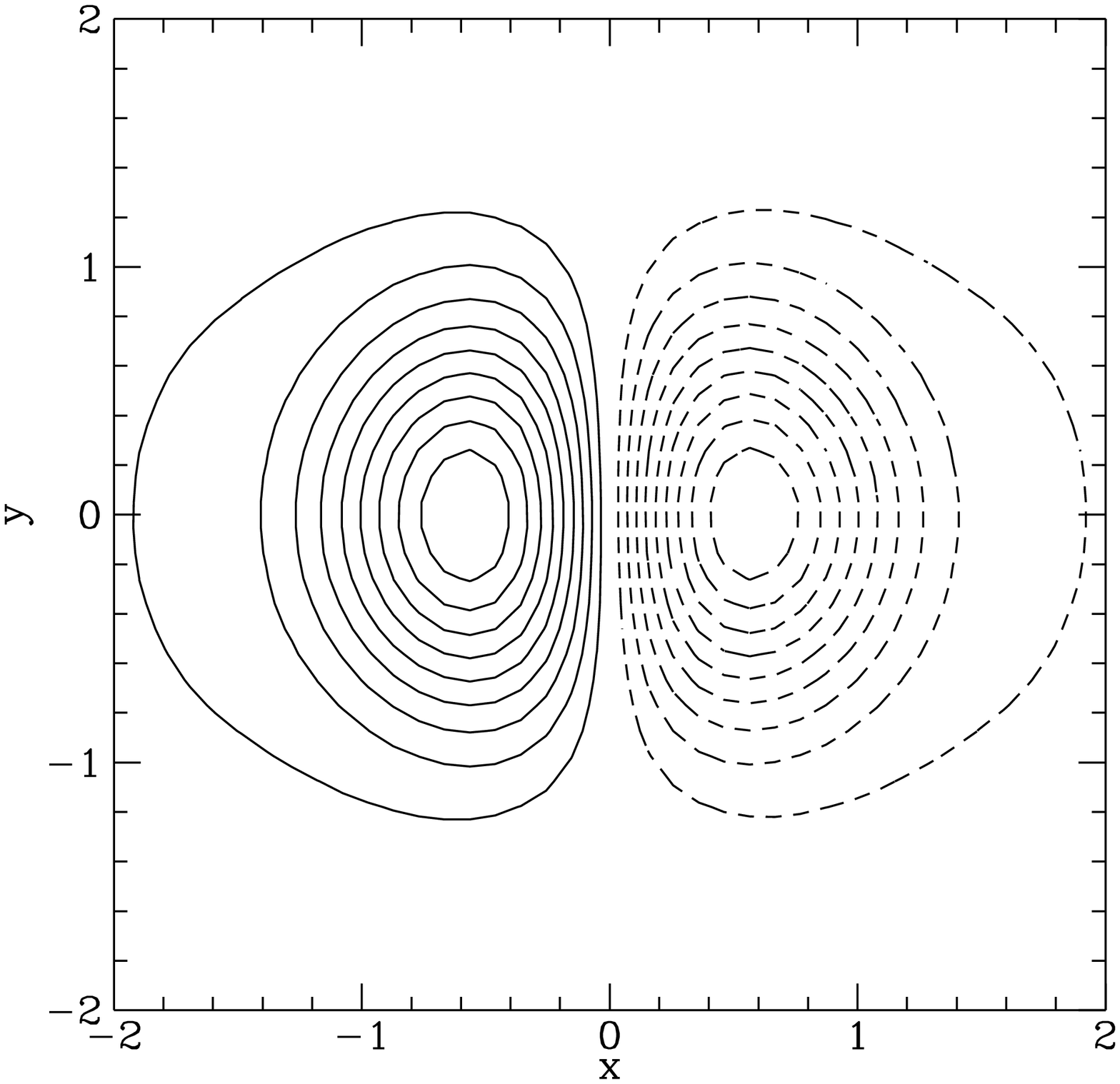}}
    \mbox{\epsfxsize=2.4in\epsfbox{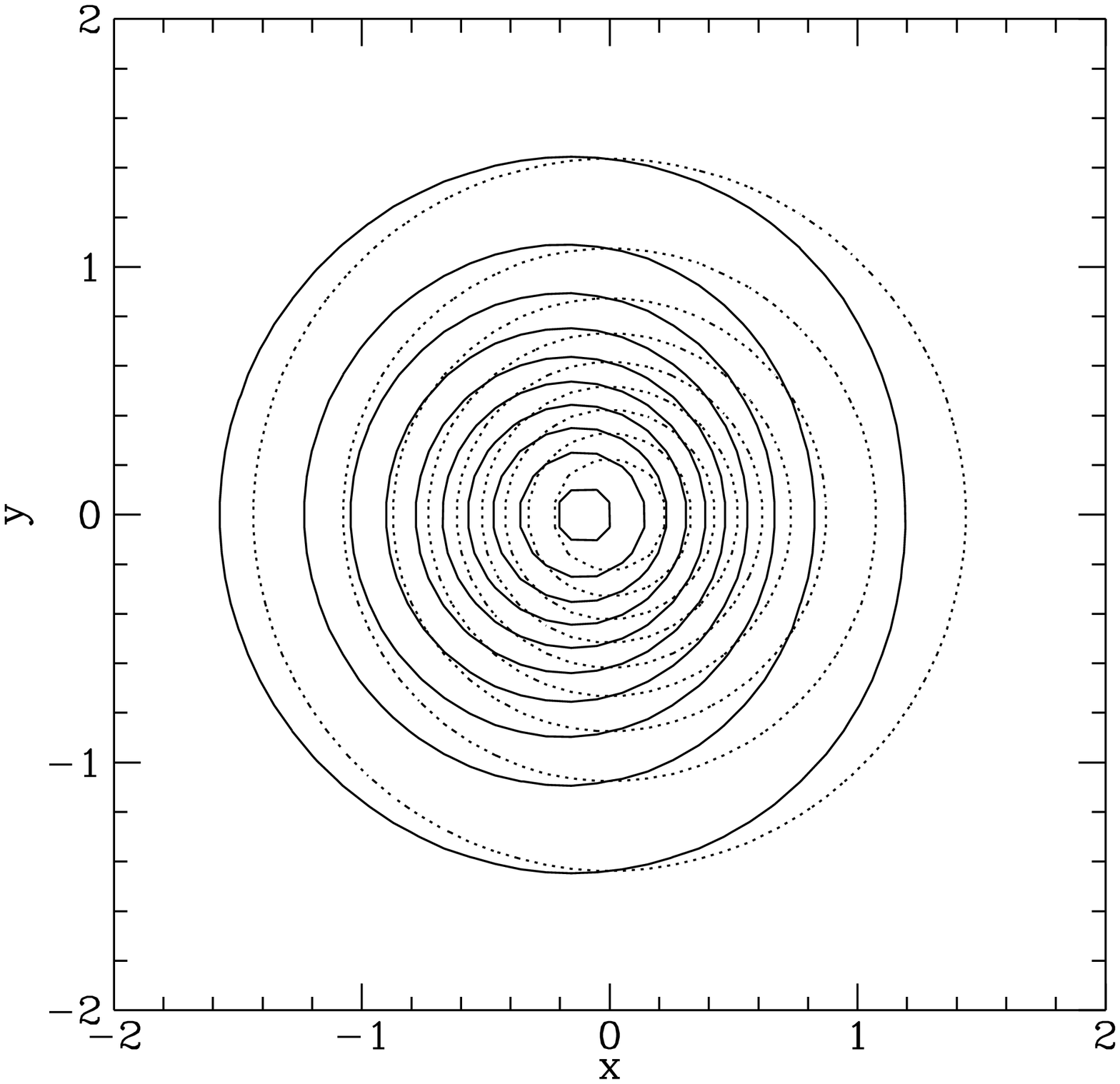}}
    }
  \caption{Weakly damped mode in $W_0=5$ King model.   Left: the
    density eigenfunction in the symmetry plane.  Overdensity
    (underdensity) is indicated by the solid (dashed) contours.
    Right: density profile of the mode in the halo (solid line)
    compared with the profile without the mode (dotted curve).}
  \label{fig:halodamp}
\end{figure}

The strongest of these halo modes is an $m=1$ or sloshing mode and
damps surprisingly slowly.  An example is shown in Figure
\ref{fig:halodamp} for a King model for clarity.  These modes exist
for all typical halo models (e.g. Hernquist models [Hernquist 1990],
NFW profiles [Navarro, Frenk \& White 1997]) and do not depend on the
existence of a core.  A qualitative explanation for the slowing
damping is as follows.  A coherent mode damps through resonances
between its pattern speed, $\Omega_p$ and the orbital frequencies,
$\Omega_r$ and $\Omega_\phi$ of individual stars.  At $l=m=1$ order,
the damping is a resonance between these frequencies of the form:
$\Omega_p - l_\phi\Omega_\phi - l_r\Omega_r = 0$, where $l_\phi \in
\{-1,0,1\}$ and $l_r \in \{-\infty,\ldots,\infty\}$.  For vanishingly
small $\Omega_p$, the only orbits near resonance are Keplerian and
therefore at outer edge of the galaxy.  If the system can have a
discrete $l=m=1$ mode at small pattern speed, the resonant orbits that can
cause damping will be at large radii and the damping will be small.
This is exactly what happens.

In the inner galaxy, many studies suggest or assume that the disk
dominates the gravitational potential.  Perhaps this reduces or
eliminates the $m=1$ halo mode?  Not so! The disk and halo can have a
mutual weakly damped $m=1$ mode.  The quadrupole mode is a weakly
amplified by self-gravity but most others appear to damp quickly.  The
$m=1$ mode is very easy to excite.  Because the halo may contain most
of the mass of the entire galaxy, a relatively small halo disturbance
can result in a large perturbation of the disk.  For example, the
Poisson noise from a halo of $10^6\msun$ black holes can produce a
mode with potentially observable consequences.  An orbiting satellite
or shot noise caused by accreting massive HVCs and dwarfs can all
potentially excite the $m=1$ mode.  Finally, we tend to think of a
galaxy as in equilibrium, however the orbits in the outer halo have
very long periods (several Gyr); therefore phase mixing is another
source of noise (Tremaine 1993).

Although I have emphasized the $l=m=1$ sloshing modes, there are also
$l=m=2$ bar-like modes.  These quadrupole responses have a shorter
damping time then the sloshing mode and this can be understood by the
same sort of argument: for a non-zero $\Omega_p$ there are more and
closer stellar orbits with resonances of the form $2\Omega_p -
2\Omega_\phi - l_r\Omega_r = 0$.   Higher order halo modes damp more
quickly. 

\begin{figure}
  \mbox{\epsfxsize=5.0in\epsfbox{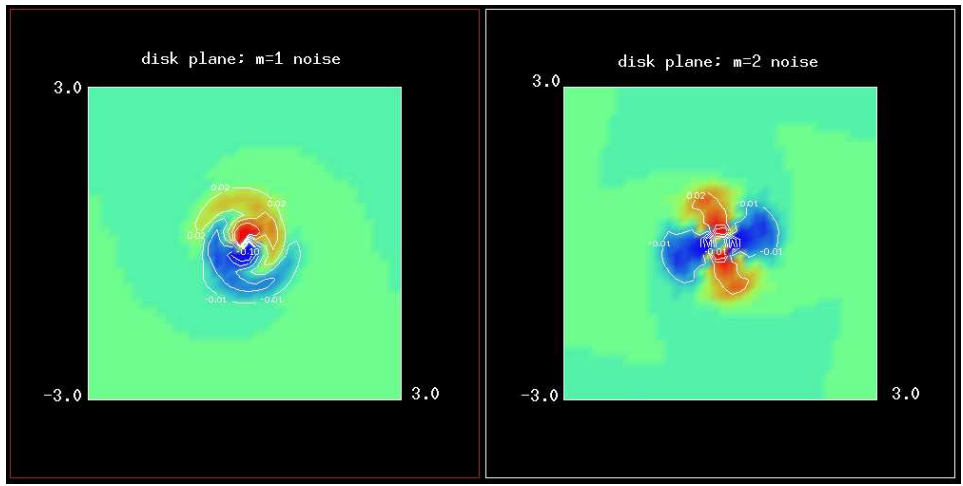}}
  \caption{Density fluctuations due to halo particle
    noise ($N=10^6$).  Amplitude of the $m=1$ response (left) is
    roughly ten times the amplitude of $m=2$ response (right).}
  \label{fig:densfluct}
\end{figure}

Why haven't these modes been seen in n-body simulations?  There are
two answers to this question.  First, they have been or at least their
effects have been.  Weakly damped modes will be excited by Poisson
fluctuations and provide excess power at their natural scales.
Analyses of n-body simulations show that this excess power is there
exactly at the location and amplitude predicted by the dynamical
theory.  Secondly, detecting the pattern itself is much harder because
the pattern speed is slow and the noise in most n-body simulations is
sufficient to wipe out the mode in a few crossing times.  This is too
short a time for the mode to set itself.  A sufficiently quiet
simulation requires many millions of particles.

As an example of the excess power, Figure \ref{fig:densfluct} shows
the effect of the modes on the response to particle (Poisson) noise.
The left-hand (right-hand) panel shows the overdensity for $l=m=1$
($l=m=2$).  The overdensity pattern is dominated by the mode.  The
$l=m=2$ profile is an order of magnitude smaller than $l=m=1$, and
both exceed that expected by Poisson fluctuations alone.  Higher order
harmonics are consistent with the expected Poisson amplitudes (see
Weinberg 1998a for additional details).  This illustrates a general
feature of halo dynamics: the response to a transient disturbance will
be dominated by the weakly damped modes, independent of the details of
the disturbance.

\subsection{Bending and warps}

\begin{figure}
  \mbox{
    \mbox{\epsfxsize=2.6in\epsfbox{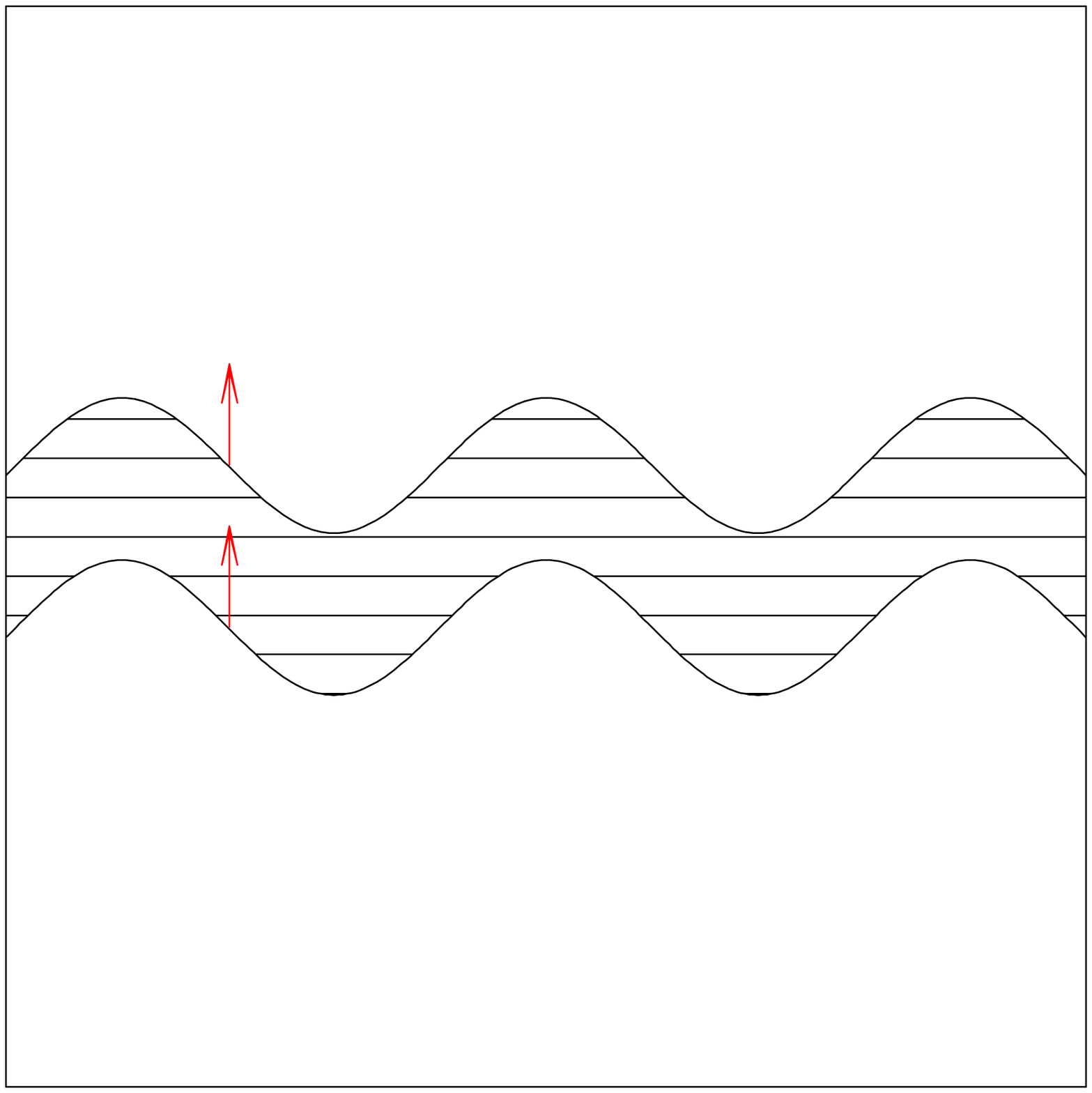}}
    \mbox{\epsfxsize=2.6in\epsfbox{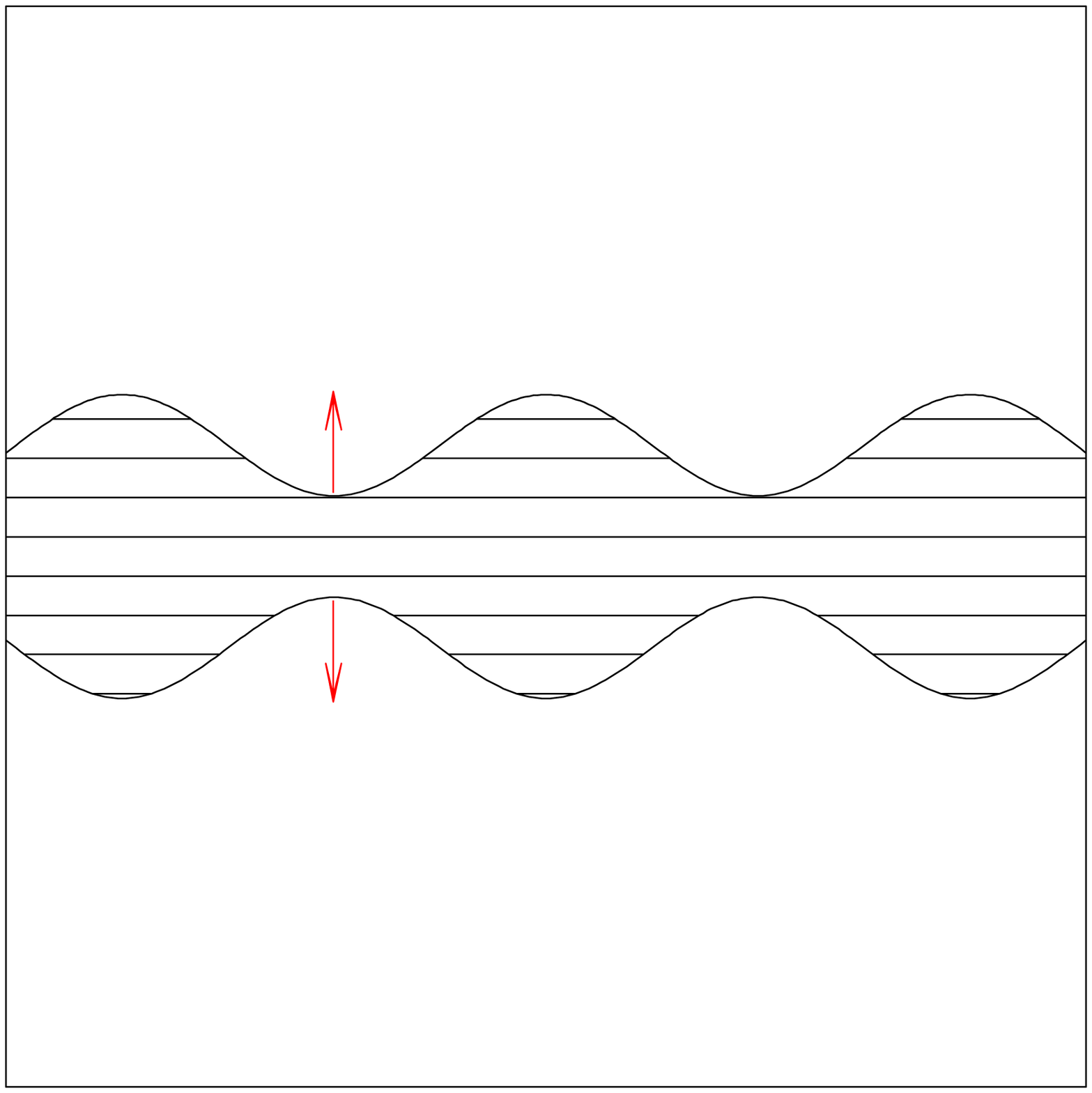}}
    }
  \caption{Bending modes in a stellar slab.  Careful solution shows
    that there are two sorts of modes, an odd or {\em bending} mode
    (left) and an even or {\em breathing} mode.  The bending mode is
    purely oscillatory in the limit that the vertical velocity
    dispersion vanishes relative to the in-plane velocity dispersion.
    The breathing mode is the damped analog to the Jeans' instability
    and approaches the unstable Jeans' mode as the velocity dispersion
    decreases.  The arrows indicate the instantaneous direction of the
    slab for each type of mode.}
  \label{fig:bending}
\end{figure}

Warps in the outer galaxy reveal some combination of the halo and disk
properties, the galactic environment and history.  This promise of
disk warping as a probe of the gravitational potential has lead to
numerous theoretical and observation campaigns (see Binney 1992 for a
review).  Whether forced by an external perturbation such as a dwarf
galaxy, accreted material or a rotating triaxial halo, warping and
bending of the disk may be considered as a response.  Just as in the
case of the halo, these features will be generic.  A good toy model is
a slab of stars of infinite horizontal extent but finite vertical
extent.  The limiting case, an infinitesimally thin stellar slab can
sustain bending modes.  In general, both bending and breathing modes
exist (see cartoon in Fig.  \ref{fig:bending}).  As one thickens the
slab, these bending modes begin to damp (Weinberg 1994).  The damping
results from a coupling between the bending and the vertical degrees
of freedom in orbits making up the disk.  If you perturb an slab with
projectile, it will begin bend.  However, the thicker the slab (ratio
of vertical to in-plane velocity dispersion), the faster the damping.

The generalization of this simple slab to an thin axisymmetric disk is
due to Hunter \& Toomre (1969) and since has been invoked to explain
disk warps (e.g Sparke \& Casertano 1988, Hofner \& Sparke 1994).
More generally, a warp may be a player in a much larger nested set of
distortions.  For example, Weinberg (1998b) suggests the possibility
of a warp created by a distortion from the orbiting LMC.  The LMC
first excites a wake in the halo.  The halo distortion will be similar
in shape to those depicted in Figures \ref{fig:halodamp} and
\ref{fig:densfluct}.  But because the LMC orbital plane is highly
inclined this wake can differentially distort the disk plane and the
disk response is a warp (more on this in \S{\ref{sec:LMC}}).  This
final topic points out that all of these distortions may be
simultaneous and connected, and at the very least, a disturbance in the
halo or disk should not be considered independently.

\section{Milky Way asymmetries}
\label{sec:MW}

Threaded through the dynamical overview is the importance of the
underlying galactic structure in forming a response, independent of the
excitation.  Before we put the whole picture together, let us take
stock of the Milky Way asymmetries.  The Milky Way shows evidence for
every type of asymmetry described above:
\begin{enumerate}
\item The Milky Way has a bar.  This has been demonstrated in many
  tracers: surface brightness (COBE), star counts, gas and stellar
  kinematics and, as originally postulated by de Vaucouleurs, based on
  morphology (de Vaucouleurs \& Pence 1978).  Recent work by
  Englmaier, Gerhard and collaborators shows this quite convincingly
  (e.g. Englmaier \& Gerhard 1999).
\item The Milky Way is lopsided.  In fact, it is lopsided both in the
  inner kpc and the outer 30 kpc!  The former asymmetry is clearly
  seen the molecular gas (e.g. Blitz et al. 1993) and the latter is
  evident in the H{\sc I} analysis presented by Henderson et al. 1984.
\item The Milky Way is most certainly warped.  This is also clearly
  seen in the H{\sc I} layer (Henderson et al. 1984).  The line of
  nodes is roughly $l=0^\circ$.  The H{\sc I} layer reaches nearly 3
  kpc above the plane at edge of the stellar disk ($R\approx 16$ kpc)
  and this is echoed in the molecular gas layer (Heyer et al. 1998).
  There is a evidence for a stellar warp as well (Djorgovky \& Sosin
  1989, Carney \& Seitzer 1993) but recent analysis based on Hipparcos
  data (Smart et al.  1998, Drimmel et al.  1999) show that some
  mysteries remain.
\item The shape of the halo is less certain.  Recent results claim
  both flattened and round.  Two arguments for a flattened halo have
  been recently presented by Olling \& Merrifield 2000.  The first is
  based on the Milky Way rotation curve and the local stellar
  kinematics.  The second assumes that the H{\sc I} is in hydrostatic
  equilibrium and uses the magnitude of gas-layer flare to estimate
  the vertical force and therefore the halo shape.  Both methods yield
  a ratio the short to long axis of $q\approx0.8$.  A recent preprint
  by Ibata et al. 2000 argue that $q\approx1.0$ based on the nearly
  great-circle appearance of carbon stars attributed to the Sgr A
  dwarf debris trail.  Any significant deviation from spherical would 
  cause the stream to precess and produce smearing which is not seen.
\end{enumerate}

\begin{table}[t]
\caption{Is the Milky Way typical?}
\label{tab:mwcompared}
\begin{tabular}{l|l}
All galaxies & Milky Way \\ \tableline
72\% barred {\small (weak and strong in H-band)}  &  Barred (SAB), 4-arm spiral\\
1.4 major satellites/spiral galaxy & Magellanic Clouds, + \\
$\sim50\%$ warped   & Warped \\
$\sim5\%$ optical, $\sim50\%$ HI lop-sided & Lopsided center, outer HI \\
Nearly all galaxies have halos (inferred)  & 10:1 \\ \tableline
\end{tabular}

\begin{tabular}{l} \small
References: \\
Eskridge et al. 2000 (bar fraction $t$-independent) \\
Zaritsky et al. 1997 \\
Reshetnikov \& Combes 1998 \\
Rudnick, Rix, Kennicut 2000
\end{tabular}

\end{table}

Is this amount of asymmetries unusual?  Or is it the case that no
galaxy is normal when you look closely?  First, anecdotally or based
on counting morphological types in the RC3, a significant fraction
(roughly 40\%) of galaxies are barred!  This turns out to be a
stronger result in near-infrared; Eskridge et al.  (2000) show that
72\% of spirals are barred (strongly or weakly).  This is expected
theoretically as the discussion in \S\ref{sec:bar} illustrates.  The
trapped orbits that make up bars must be dynamically old and therefore
relatively red compared to colors of populations with recent star
formation.  Second, all galaxies have dwarf companions that may
trigger asymmetries. In an ambitious project designed to measure the
mass of dark halos, Zaritsky and collaborators observed dwarfs around
isolated field galaxies.  Their census shows that the average number
of companions of the LMC sort are 1.4 (Zaritsky et al.  1997).  The
LMC luminosity is a ``one sigma'' on the bright side of the mode.
Similarly there are both closer and more distant dwarfs.  In short, we
can expect most spirals to have a nearby dwarf companion.  Third, what
about warps?  Statistics on warps are more difficult to obtain but
roughly half of spirals have outer warps (Binney 1992, Burton 1998)
with later types predominating (Bosma 1991). Anecdotally, people who
look hard at morphology claim that all spiral galaxies are warped.
Fourth, more than half of spirals appear to be lopsided in H{\sc I}
(Swaters et al.  1999) and 20\% of all disk galaxies (Rudnick \& Rix
1998) and a recent paper by Rudnick, Rix \& Kennicut (2000) estimate
that 5\% are lopsided in the optical.

Table \ref{tab:mwcompared} shows summarizes these facts.  In short,
all galaxies seem to show very similar sorts of asymmetries.  Because
the halo dominates the mass of a galaxy we expect that any
non-axisymmetric structure in a halo will be manifest in the
lower-mass luminous component.  The strongest easy-to-excite feature,
the $m=1$ mode, can cause both lopsidedness and warps as I am about to
describe.

\section{Satellites and dwarfs}

\subsection{The LMC}
\label{sec:LMC}

\begin{figure}
  \mbox{\epsfxsize=5.0in\epsfbox{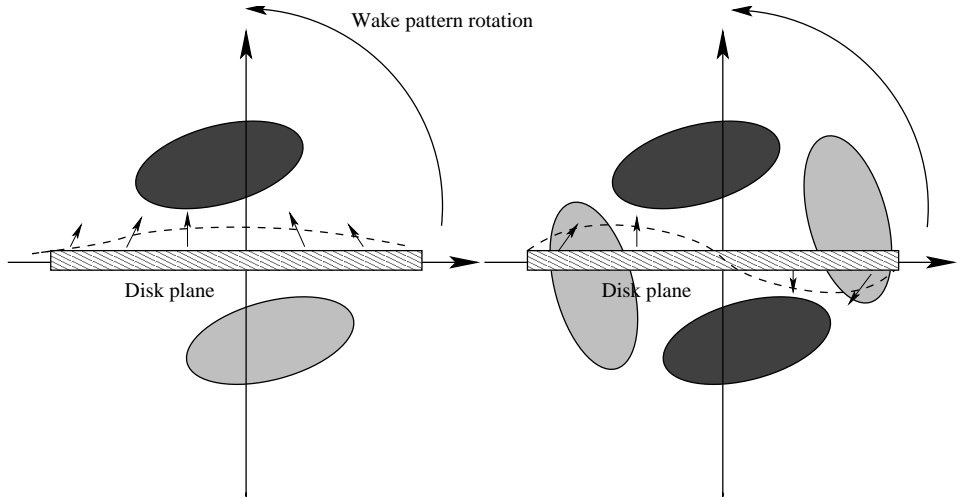}}
  \caption{Sketch showing the differential force on a disk from
    a dipole wake (left) and quadrupole (wake) right.  The dark and
    light grey ovals indicate overdensity and underdensity,
    respectively.  In the dipole ($l=m=1$) case, the force is one
    sided, giving rise to a dish-shaped distortion.  In the quadrupole
    ($l=m=2$) case, the force is bisymmetric giving rise to an
    integral-sign shaped distortion.  The dashed lines indicate the
    expected shape of the warp in each case.}
  \label{fig:satwarp}
\end{figure}

Hunter \& Toomre (1969) explored the possibility of the LMC being the
source of the Galactic warp and found that a direct tidal perturbation
only resulted in a warp with an amplitude of 50 pc!  Of course, this
predated the dark matter paradigm and the halo was not included.  A
major part of the same wake that causes LMC to decay by dynamical
friction is the dipole and quadrupole distortions, amplified by the
self-gravitating mode discussed earlier\footnote{In fact, one can show
  rigorously that the response of the halo, the wake, attracts the
  satellite that excited it with exactly the force predicted by
  dynamical friction theory which then causes the satellite orbit to
  decay.}.  Because the LMC is on a polar orbit the response is not in
the disk plane and the wake will exert a differential vertical force
on the disk.  In particular, the quadrupole distortion can cause an
integral-sign shaped warp (see cartoon in Fig.  \ref{fig:satwarp} for
an explanation).  The dominant effect of the dipole will be a center
of mass shift.  The warp is strongest if the pattern speed of the warp
is close to the pattern speed of the halo wake.  It turns out that the
LMC is nearly in an ideal position for this to happen.  In addition,
the dipole part ($m=1$) causes a significant in-plane asymmetry in the
outer part of the halo quite similar to the Henderson et al. (1984)
figure discussed earlier.

Some recent work (Garcia-Ruiz, Kuijken \& Dubinski 2000) has suggested
that the LMC can not be the source of warp because the response will
be in phase with the satellite.  A detailed study of the response
reveals that the mean position angle of the outer wake is roughly
aligned but, the inner wake trails considerably (greater than
$45^\circ$ at various points in the orbit).  Physically, this phase
lag must occur in order for there to be any dynamical friction; it is
the attraction by the wake the causes the drag.  The halo wake is not
just an added enhancement to the satellite torque but dominates the
direct satellite perturbation by nearly an order of magnitude inside
of half the LMC orbital radius.  This inner wake is dominated by a
higher-order resonances (the 2:1 resonance is often strong).  It is
likely that the parent orbits for such a resonance could exchange
stability and appear aligned rather than perpendicular.  To summarize,
the amplitude and morphology of the wake will depend on details
galactic potential and the satellite orbit, but conversely is not
strongly constrained by orbital positional angle.  Although a strong
warp requires a nearby resonance, this obtains for a broad range of
conditions.

\begin{figure}
  \mbox{
    \mbox{\epsfxsize=2.6in\epsfbox{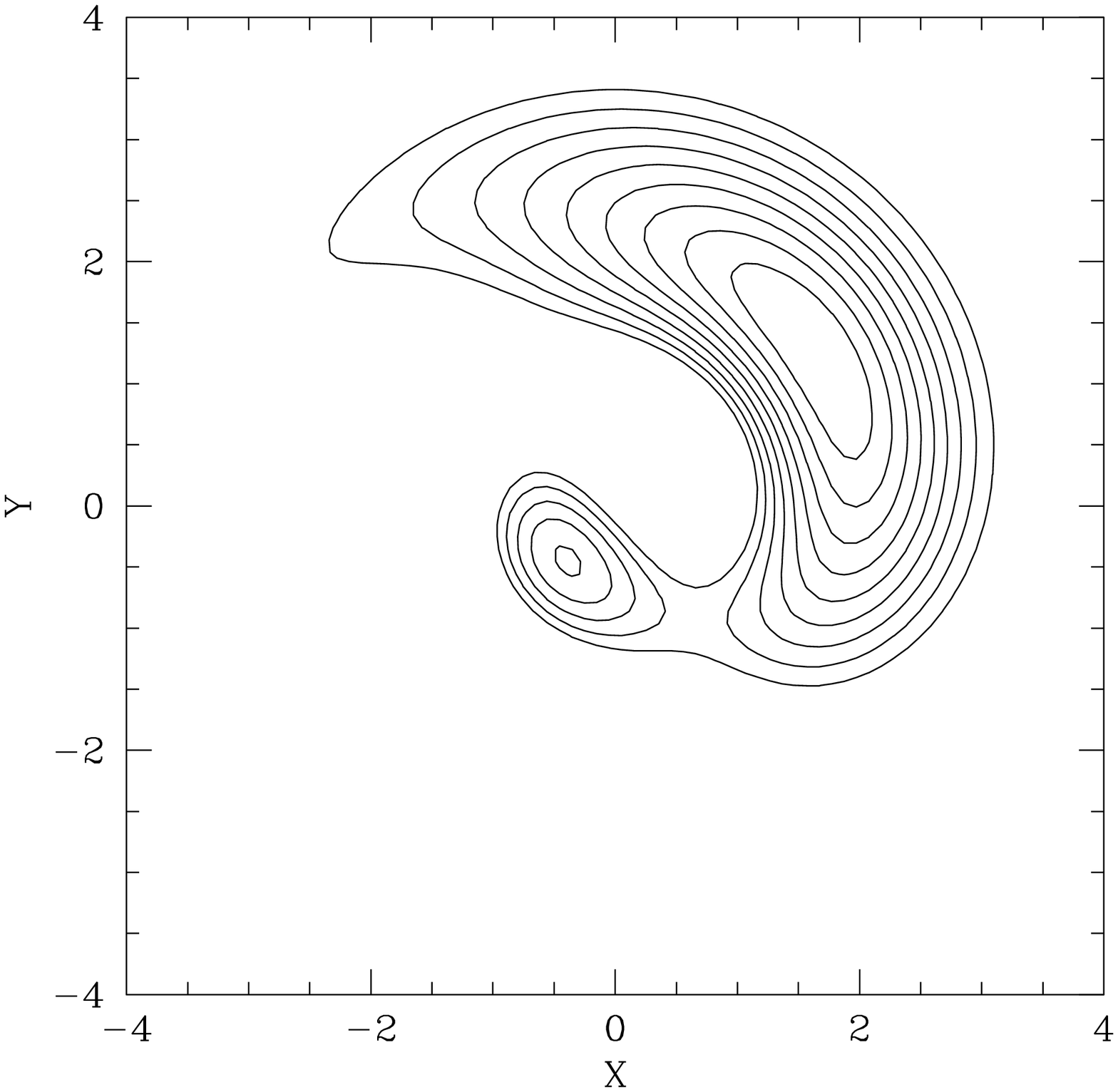}}
    \mbox{\epsfxsize=2.6in\epsfbox{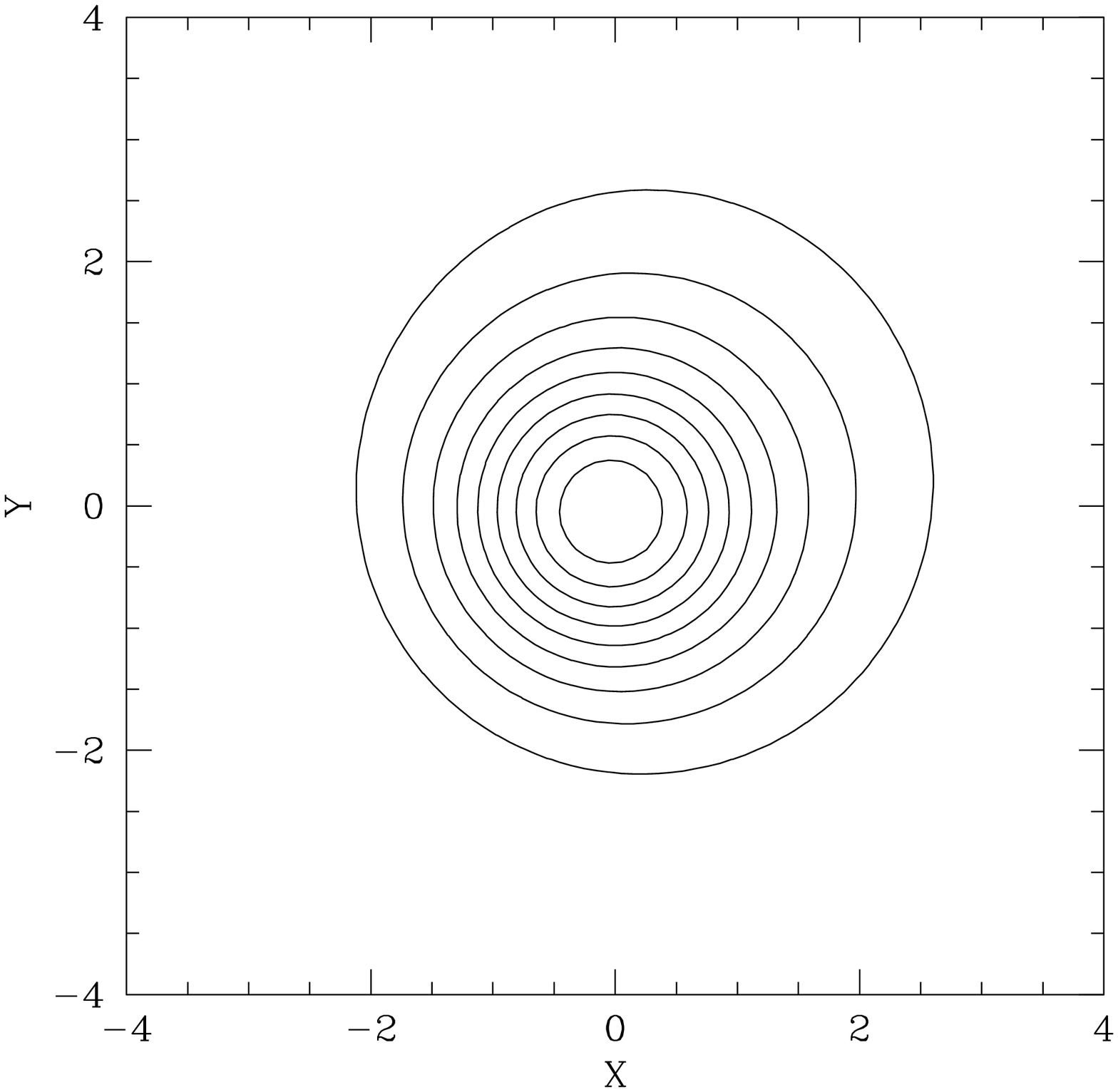}}
    }
  \caption{Left: shape of $m=1$ in-plane distortion for a
    $2\times10^{10}\msun$ LMC mass.  Right: the isodensity profiles
    for this distortion Note that the inner galaxy is shifted in the
    opposite sense as the outer galaxy, however, unless there is a way
    of inferring the unperturbed center, this would not be
    measurable.}
  \label{fig:inplane}
\end{figure}

Scalo (1987) argued the possibility that the Milky Way and the LMC
have time-synchronous bursts and this is consistent with the
chromospheric age work by Barry (1988).  Although it is easy for the
LMC to be strongly affected by the tidal field of the Milky Way
(Weinberg 2000), it is more difficult for the LMC to affect the Milky
Way.  This is only a suggestion, but note that resonant excitation of
the halo modes is a plausible mechanism.

Just as interesting is the effect of the Milky Way on the local group
dwarfs.  After working hard to understand the LMC-induced Galactic
structure, I have recently been impressed with the damage that may be
done to a dwarf by the Milky Way.  Dwarfs within roughly 100 kpc are
likely to be tidally limited.  By definition, the gravitational force
of the Milky Way on mass at the tidal radius is of the same order as
the gravitational force exerted by the dwarf.  At half the tidal radius,
the tidal force is roughly an order of magnitude smaller but a 10\%
perturbation is still significant and leads to a number of interesting
responses:
\begin{enumerate}
\item Disk heating: the dependent forcing as the dwarf moves on its
  orbit can heat the disk noticibly in several gigayears.
\item Precession: the disk plane is unlikely to be coplanar with the
  orbital plane, the resulting torque will cause the disk to precess.
  The leads to a complex interaction between the global torque, halo +
  disk back reaction and resonant heating.
\end{enumerate}
This complex interaction should lead to disk warping and other
differential distortions.  This motivates a increasingly detailed
study of the LMC-Milky Way interaction which may help us appreciate a
suite of more general problems.  For example, will this complex of
dynamical mechanisms destroy a disk to make spheroidal?  Will the
heating exacerbate loss of equilibrium in the Galactic tidal field
leading to stellar and gaseous streams?  More quantitatively, can we
predict the disruption rate?

\subsection{Dwarf fly-by events}
\label{sec:flyby}

\begin{figure}
  \mbox{\epsfxsize=\textwidth\epsfbox{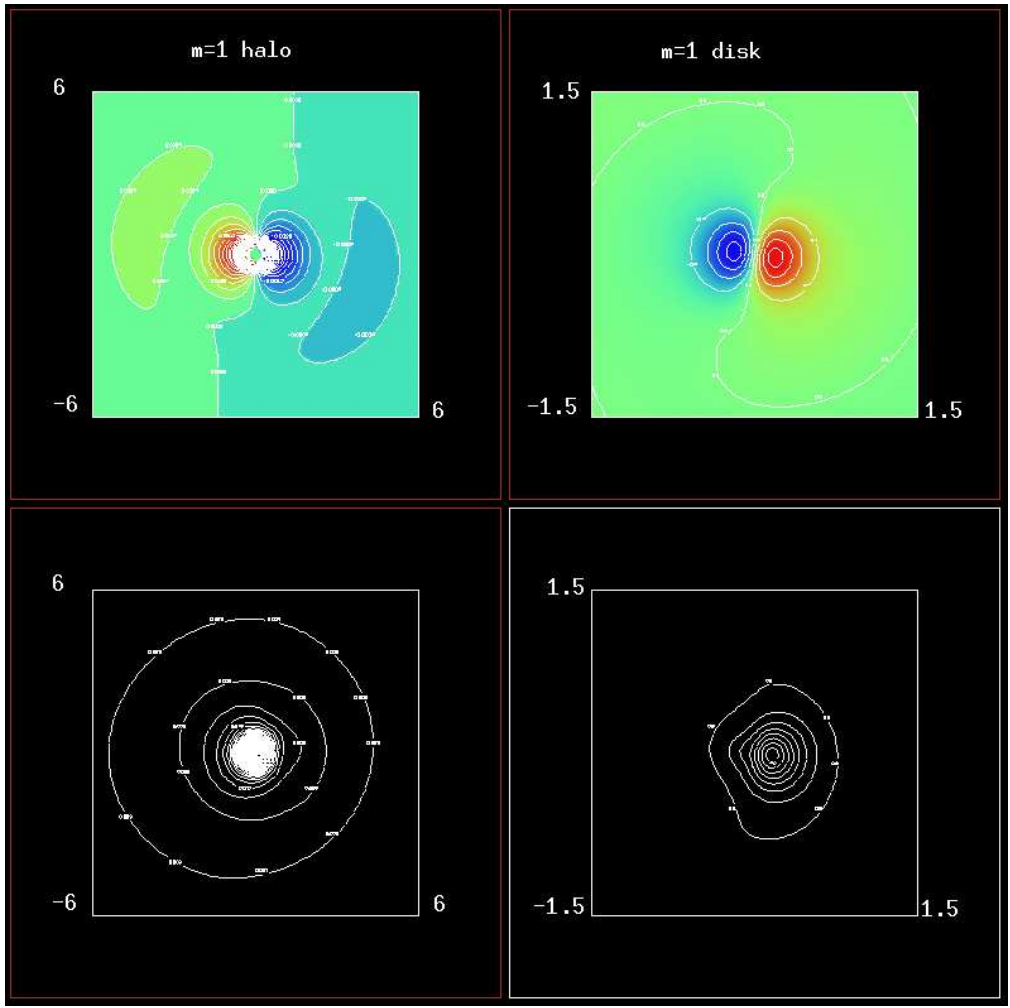}}
  \caption{$m=1$ halo (upper left) and
    disk (upper right) response to a perturber flying by on the disk
    plane. Lower panels show the total halo (left) and disk (right)
    density on the $x-y$ plane (from Vesperini \& Weinberg 2000b).
    Notice that the disk and halo responses are anti-correlated.}
  \label{fig:nbodyflyby}
\end{figure}

\begin{figure}[tbh]
  \mbox{
    \epsfxsize=2.6in\epsfbox{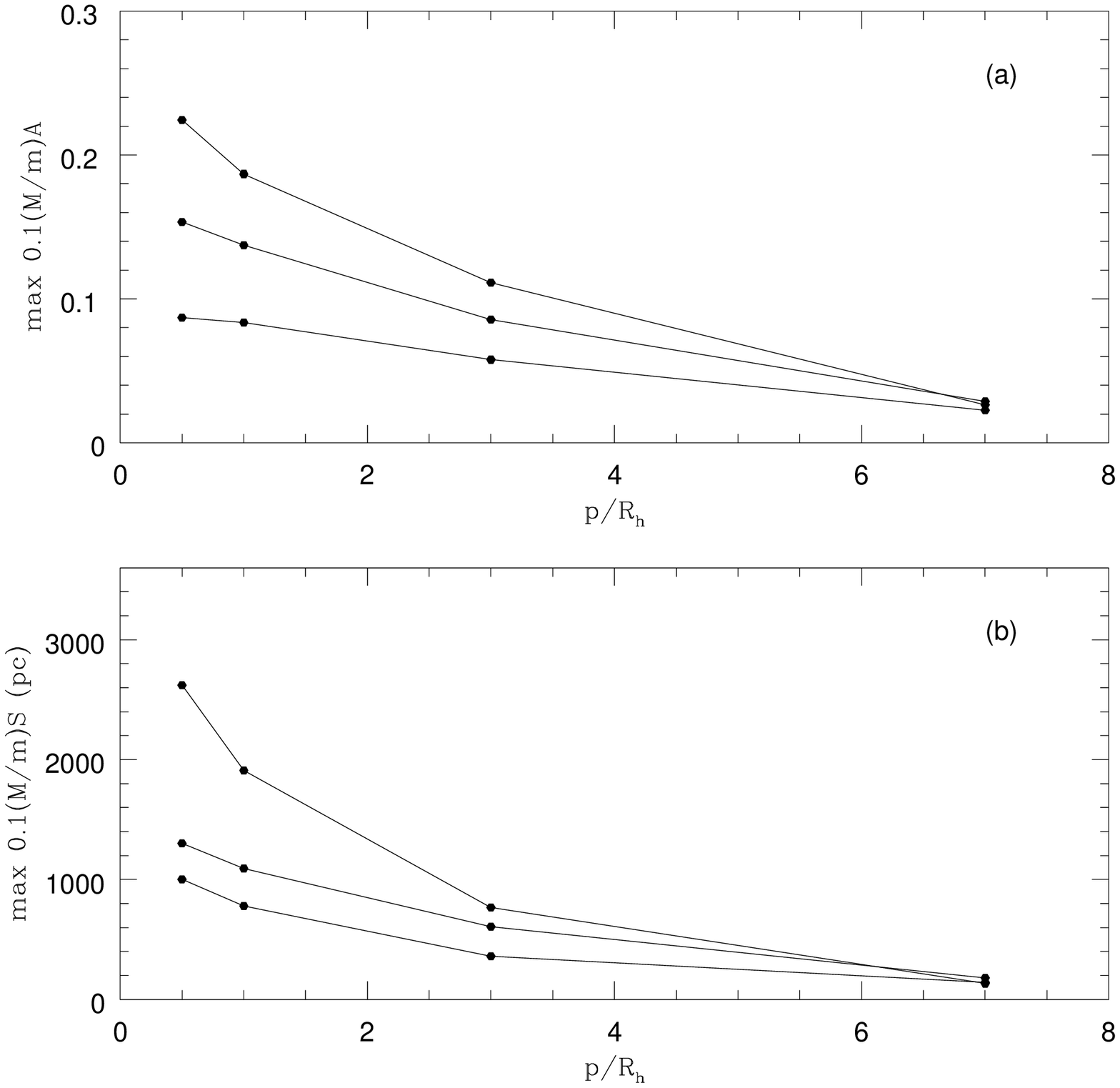}
    \epsfxsize=2.6in\epsfbox{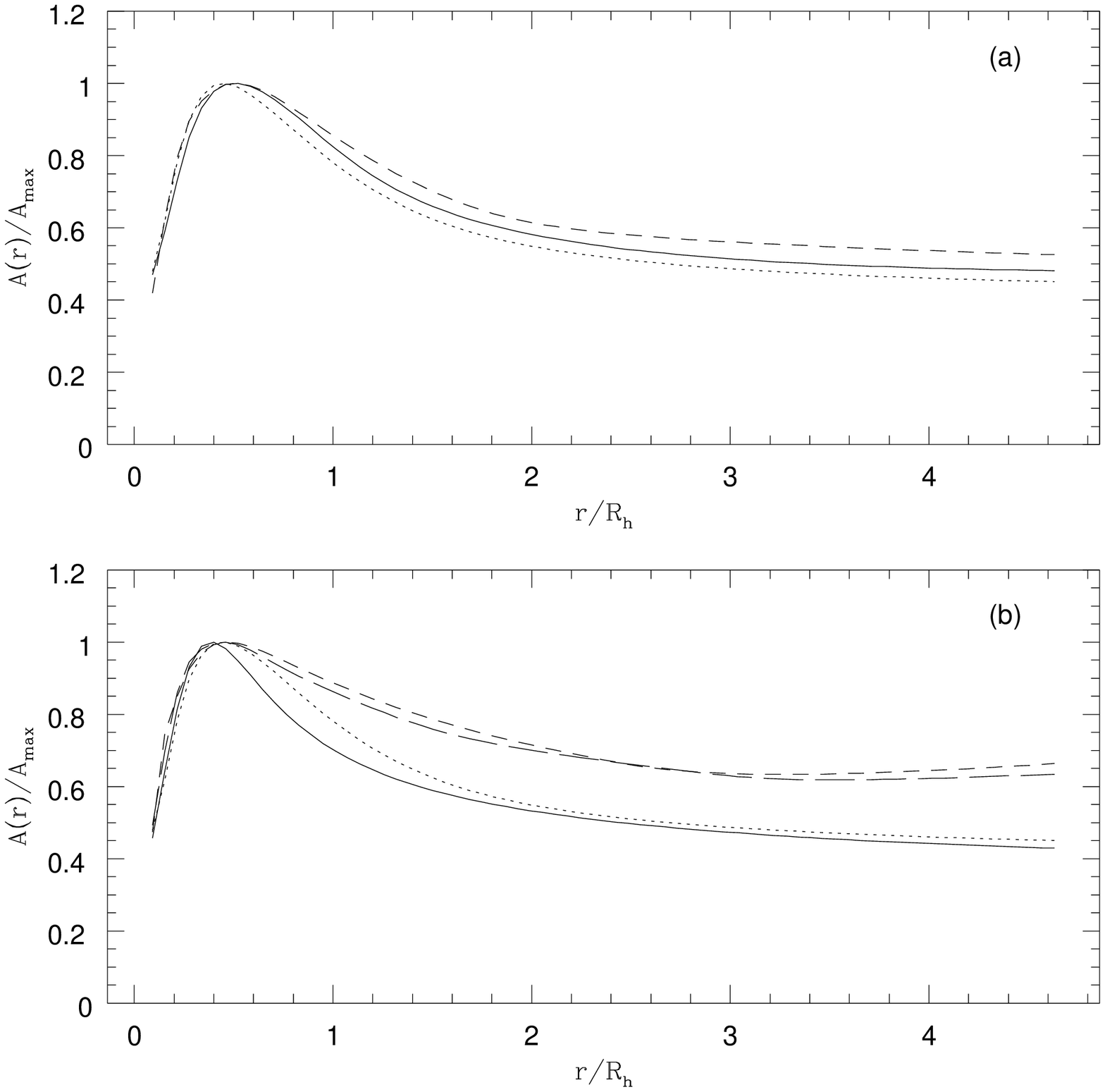}
    }
  \caption{Top left: Variation of asymmetry parameter $A$ with
    pericenter $p/R_h$ for perturbers with 10\% of the galactic halo
    mass (scaling shown). The three curves correspond (from the upper
    to the lower curve) to $V=200,~500,~1000$ km/s.
    Bottom left: Center-of-mass offset in
    parsecs for the same events. The three curves correspond (from the upper
    to the lower curve) to $V=200,~500,~1000$ km/s.
    Top right: Radial profile of $A$ normalized
    to its maximum value for fly-bys with $p/R_h=1.0$ and $V=200$ km/s
    (solid line), $V=500$ km/s (dotted line), $V=1000$ km/s (dashed
    line); each curve correspond to the radial profile of $A$ when the
    total $A$ is maximum. 
    Bottom right: Same as top right but for fly-by with $V=500$
    km/s and $p/R_h=0.5$ (solid line), $p/R_h=1.0$ (dotted line),
    $p/R_h=3.0$ (dashed line), $p/R_h=7.0$ (long dashed line).  }
  \label{fig:asymm}
\end{figure}

\begin{figure}[tbh]
  \centering
  \mbox{\epsfxsize=3.5in\epsfbox{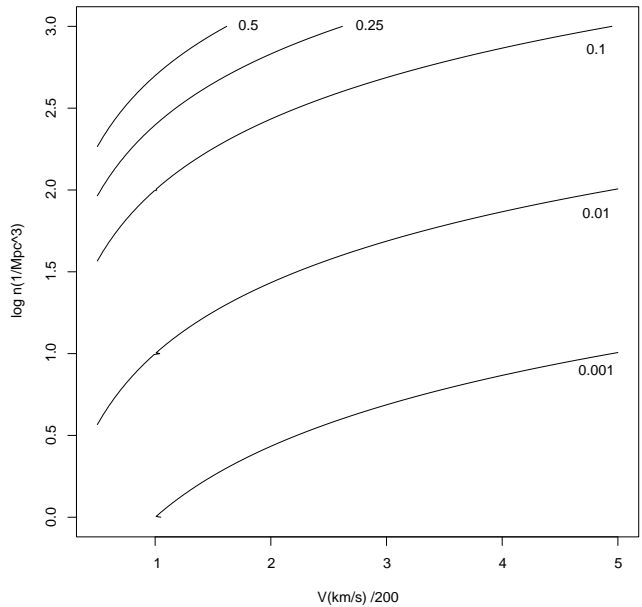}}
  \caption{Probability of $A\ge0.1$ in the last $10 t_{dyn}(R_h)$.  
    The vertical and
    horizontal axes describe the number density of the environment and
    relative velocity of encounters.}
\label{fig:aprob1}
\end{figure}

\begin{figure}[tbh]
  \centering
  \mbox{\epsfxsize=3.5in\epsfbox{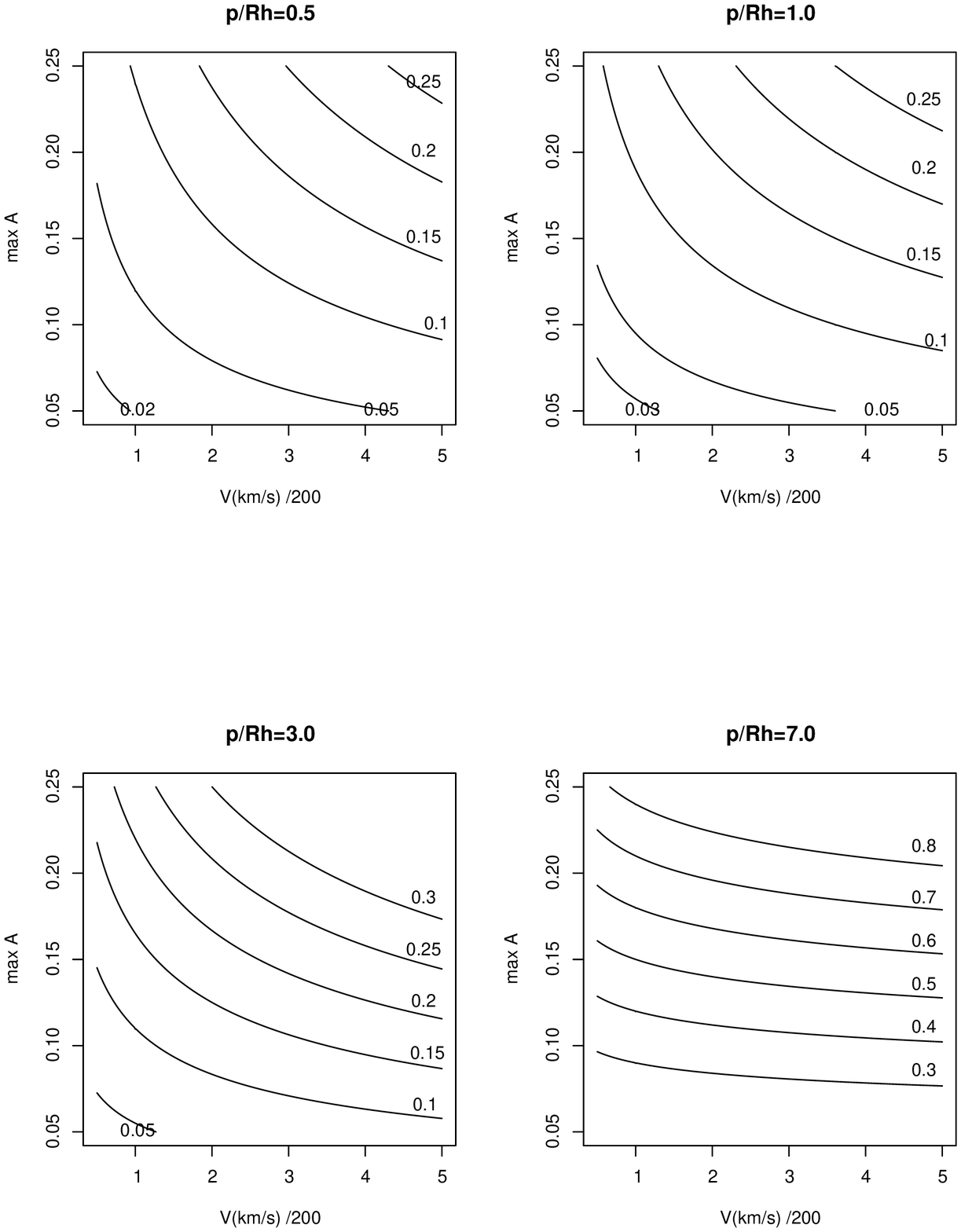}}
  \caption{Relative mass (in halo mass units) for the given pericenter
    to halo half-mass radius ($p/R_h$) to to produce $\max A$ for the
    encounter velocity $V$.}
\label{fig:aprob2}
\end{figure}

Another way of getting the same sort of excitation, perhaps more
important for group galaxies than the Milky Way, is a passing fly by.
A perturber on a parabolic or hyperbolic trajectory can excite similar
sorts of halo asymmetries and persist until long after the perturber's
existence is unremarkable.  Presumably, our Galaxy has suffered such
events in the past but because the satellite excitation is closely
related to the fly-by excitation, the study of one will provide
insight into the other. The upper left and right panels of  Figure
\ref{fig:nbodyflyby} show the 
non-axisymmetric $m=1$ part of the disk and halo perturbation from an
n-body simulation (Vesperini \& Weinberg 2000b).  The
perturber trajectory is on the $x$--$y$ plane with pericenter at 70 kpc
and mass twice that of the LMC. The lower left and right panels of
Figure \ref{fig:nbodyflyby}  
show the resulting halo and disk distortion. The distortion in the disk
plane closely follows that in the halo.  This is example has a
dramatic distortion but smaller effects would be easily detectable.

The n-body simulations agree in magnitude and morphology with the
perturbation theory.  The perturbation theory allows us to scale the
encounter by velocity and impact parameter (or their ratio if the
perturber is completely outside of the halo).  From the analytic
calculations, we can compute the standard asymmetry parameters
(e.g. Abraham 1996a, Conselice et al. 2000) obtained by summing over
the mean square difference of the galaxy and its $180^\circ$ rotated
image:
\begin{equation}
  A = {1\over2} {\sum\left|I(x,y) - I_{rot}(x,y)\right|\over \sum I(x,y)}.
  \label{eq:Adef}
\end{equation}
The results are shown in Figure \ref{fig:asymm} (upper left) which
shows the maximum value of $A$ for an impact parameter $p$ in units of
the half-mass radius $R_h$ for three different encounter velocities:
$V=200$ km/s, $V=500$ km/s and $1000$ km/s.  
The amplitude, of course, depends on the perturber mass and is shown
here for 10\% of the halo mass but this can be scaled to any desired
value. For this figure the model adopted for the halo is a $W_0=7$
King model.  
For example, a perturber with 10\% of the halo mass
(approximately 3 of an LMC) with pericenter at the halo half-mass
radius and encounter velocity of 200 km/s will produce $A\approx 0.2$.
The lower left panel predicts the displacement between the position of
the peak of the density and the position of the center of mass, $S$.
For the same example, we have $S\approx2\kpc$.  Of course, this is
relative to the unperturbed position and ``observed'' figure will be a
lopsided halo as in Figure \ref{fig:inplane} (right).  Because the
halo response is dominated by the modes of the halo rather than
properties of the perturber, we expect that the asymmetry should be
dominated by contributions at well-defined radii, independent of the
perturber parameters.  We propose a simple generalization of equation
(\ref{eq:Adef}) to test this prediction: define $A(r)$ to be the sums
over pixels restricted to those within projected radius $r$.  The
results for wake excited asymmetries are shown in the right-hand-side
panels.  For a variety of encounter velocities and impact parameters,
the profiles of $A(r)$ are quite similar in shape.  n-body simulations
of flyby encounters (Vesperini \& Weinberg 2000b) have revealed that
the radial profile of $A$ for a disk has a  similar shape. 

Overall, the values of $A$ caused by fly-bys range from near zero to
$A \approx 0.2$.  This coincides with range of $A$ for galaxies in the
Medium Deep Survey and in the Hubble Deep Field (Abraham et al. 1996a,
1996b).  For a fly-by event, the $A$ parameter reaches its maximum
values when the perturber is close to the pericenter of its orbit but
these distortions are sufficiently long-lived that large values of $A$
may be observed without any close companion in the vicinity of the
primary.

Figures \ref{fig:aprob1} and \ref{fig:aprob2} quantifies the amplitude
of $A$ for in various scenarios.  Figure \ref{fig:aprob1} shows curves
constant probability for an encounter in the past $10t_{dyn}(R_h)$
(where $t_{dyn}(R_h)\def \pi(R_h^3/2GM)^{1/2}$ is the dynamical time
at the half-mass radius) such that maximum value of $A$ is greater
than 0.1.  Since these results are for spherical systems, they apply
to either the dark halos of spiral galaxies or elliptical galaxies.
Figure \ref{fig:aprob2} provides different look at the same problem.
It shows curves constant relative perturber mass necessary to have a
maximum value of the asymmetry parameter equal to $A$ during an
encounter with relative velocity $V$ for a variety of different impact
parameters (in units of halo-half mass radii $p/R_h$).  Predictions
for spiral disks are in progress.  In short, a fly-by encounter can
result in the observed values for $A$ and persist for several Gyr (see
Vesperini \& Weinberg 2000a).

\section{Summary}

In summary, I have attempted to explain how and why both satellite
interactions and fly-by encounters can produce large-scale distortions
in galaxies of the type commonly observed.  Key is the understanding
that the shape of large-scale halo and disk modes will be similar
regardless of the perturbation source.  This is intuitively familiar, I
think, from work on spiral density waves from the 1970s.  The new
twist emphasized here an analogous dynamical mechanism also leads to
global halo modes.  Since the halo dominates the mass of the galaxy,
it is intimately coupled to the morphology of the luminous disk and
spheroid.  For example, distortions in any of the components can not
be considered independently because all components respond to each
other.  In particular, satellites, fly-by encounters, and other group
interactions can effect a disk by transmission via the halo.

Because the response of a galaxy to a disturbance depends more on the
properties of a galaxy than the details of the disturbance, the
dynamical mechanisms discussed here predict that galaxy morphology
tends to be convergent.  In other words there are most likely many
paths to similar dynamical response.  These considerations lead back
to the title: the Milky Way appears to be a typical galaxy showing all
manner of asymmetries---lopsidedness, bars, arms, warps---typical of
spirals and discussed here as expected from the halo--disk
interaction.  A detailed theoretical understanding of these
interactions will therefore required detailed observations that can
only be explored in the Milky Way.

These ideas lead to the possibility that global modes are continuously
excited by a wide variety of events such as disrupting dwarfs on
decaying orbits, infall of massive high velocity clouds, disk
instability and swing amplification and the continuing equilibration of
the outer galaxy.  The dominant halo modes are low frequency and low
harmonic order and therefore can be driven by a wide variety of
transient noise sources.  Some recent work (Weinberg 2000ab) provides
a theory excitation by noise and applies this to the evolution of
halos.  Preliminary results suggest that noise may drive  halos
toward approximately self-similar profiles.  Additional work will be
required to make precise predictions for these trends and explore the
consequences for long-term evolution of disks in spiral systems.

Of course, this whole scenario of halo-mediated and halo-driven
structure is predicated on the existence of a gravitationally
interacting dark halo as opposed to some type of modified gravity.
Despite recent papers advancing MOND as a solution as a solution to
several theoretical conundrums (e.g. Brada \& Milgrom 1999ab, 2000ab),
I have not seen a satisfactory description of an evolving Universe
without dynamical friction.  Dynamical friction an essential
ingredient for hierarchical formation through galaxy merging, and in
producing remnant morphology, such as tidal tails.  This is quite
naturally and consistently explained by dynamical friction which
requires gravitational dark matter.

\section{Future prospects}

How does one test this idea that a lively halo is major player to
galaxy structure? 

\subsection{Photometry}

\begin{figure}[thb]
\centering
\mbox{\epsfxsize=3.25in\epsfbox{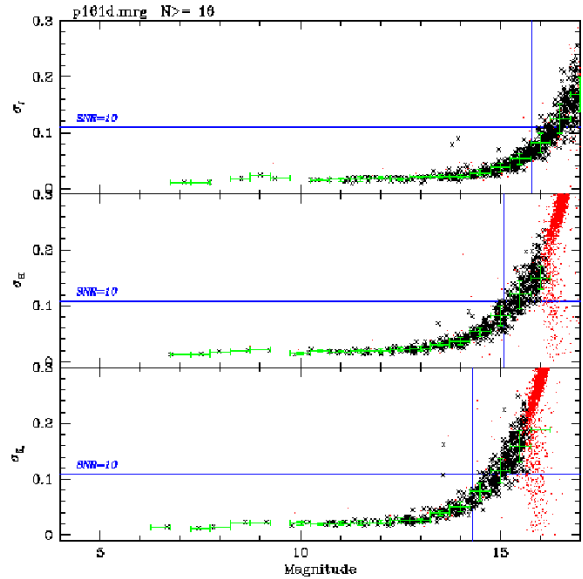}}
\caption{Photometric error estimates for 2MASS points sources in a
  typical calibration observation.  Data for individual sources
  plotted along with mean and standard error in half-magnitude bins.
  The crosses are detections in most of the observed frames (reliable)
  and the dots are detections in only a small number of frames
  (unreliable). The survey specification is S/N=10 (horizontal lines)
  at the limits given in the second column of Table
  \protect{\ref{tab:2mass}} (vertical lines).  Note that the crossing
  point varies with seeing and airglow.  In general, the survey has
  exceeded its target.  For high $S/N$, the standard relative error in
  each band less than 0.03 mag.}
\label{fig:2mass}
\end{figure}

2MASS ({\tt http://pegasus.astro.umass.edu}) will catalog
approximately 400 million stars ($S/N=10$, cf.  Fig. \ref{fig:2mass})
with over 1 billion detections (see Table \ref{tab:2mass}).  As of
this writing (September 2000) 2MASS has achieved complete full-sky
coverage.  The intrinsic low-extinction in the near infrared will
allow us to probe the Milky Way structure using star counts.  Giants,
carbon stars and other AGBs are ideal candidates for kinematic
follow-up and can be selected based on near-infrared colors.  The
underlying stellar structure is better represented in the
near-infrared than optical bands as discussed in \S\ref{sec:MW} for
barred galaxies.  In addition to direct photometric detection of
asymmetries in both stars and gas, microlensing is already confronting
our understanding of inner Galaxy morphology and the dynamics of the
LMC.  Finally, we can compare with external galaxies.  2MASS galaxy
photometry should permit a thorough exploration of asymmetry in the
near infrared using diagnostics described in \S\ref{sec:flyby} as well
as more standard Fourier diagnostics.

\begin{table}[ht]
\caption{2MASS point source sensitivity (mag)}
\begin{tabular}{l|l|l}
  Band & Completeness\tablenotemark{a}& Detection\\ \hline
  J & 15.8 & 17.3\\
  H & 15.1 & 16.6 \\
  K$_s$ & 14.3 & 15.8\\
\end{tabular}
\tablenotetext{a}{at the level $P=0.9995$}
\label{tab:2mass}
\end{table}

\subsection{Kinematics}

\begin{figure}  
\centering
\mbox{\epsfxsize=3.25in\epsfbox{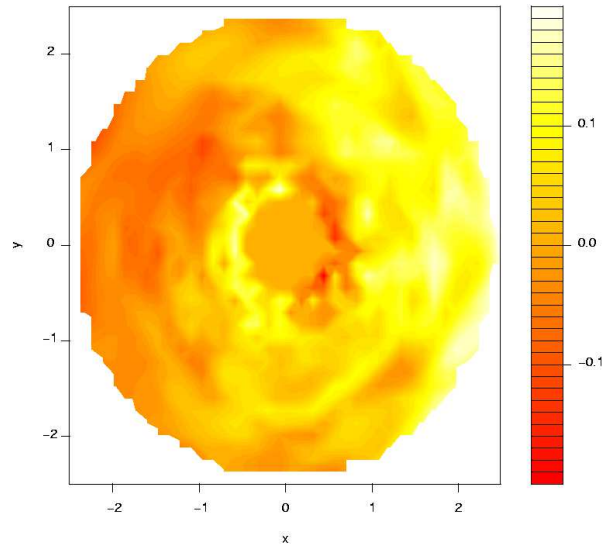}}
\caption{Tangential kinematic signature in disk induced by a fly-by
  encounter with twice the LMC mass. One length unit is 7 kpc.  The
  center of the galaxy is removed for clarity.  The asymmetry in the
  relative tangential velocity field,
  $[v_\phi(r,\phi)-v_c(r)]/v_c(r)$, is color coded (see key at
  right).}
\label{fig:nbodykine}
\end{figure}

Most important, I believe, will be prediction of detailed photometric
and kinematic signatures.  Such a joint inference has been quite
important in the last 5--10 years in establishing the existence of the
Milky Way bar.  In addition to direct radial velocity campaigns,
astrometry promises to revolutionize the study of Galactic kinematics.
A number of major astrometric missions are in progress or being
planned.

The wakes described in the sections above must induce mean flows and
we can use n-body simulations of satellite--halo interaction to
investigate globally correlated kinematic signatures.  For example,
Figure \ref{fig:nbodykine} shows the deviations from the average
tangential velocity for a disk perturbed by a perturber with a mass
equal to twice the LMC mass during a fly-by with pericenter at 70 kpc
(Vesperini \& Weinberg 2000b).  We used twice the LMC mass to increase
the signal to noise but expect linear scaling with perturber mass.
The plot shows the kinematic distortions after the pericenter passage
when the perturber is about 200 kpc away from the center of the
primary galaxy.  The interaction produces mean velocity peaks of 30
km/s in a clear dipole pattern.  Similarly, because stars do populate
the halo regardless of their origin, we can use halo stars to trace
wakes directly.  We expect the LMC to produce flows a factor of two
smaller.  It should be possible to see such a signature at
galactrocentric radii of 16 kpc with 20$\mu$as precision astrometry
(one year baseline).  This would requires a directed l.o.s. attack
with SIM or the full-survey capabilities of GAIA.
    
\acknowledgments I thank Enrico Vesperini and Mike Skrutskie for
comments on and help preparing this manuscript.  This work was
supported in part by NSF AST-9529328.  This publication makes use of
data products from the Two Micron All Sky Survey, which is a joint
project of the University of Massachusetts and the Infrared Processing
and Analysis Center, funded by the National Aeronautics and Space
Administration and the National Science Foundation.


\begin{references}
  
  \reference Abraham, R.\ G., van den Bergh, S., Glazebrook, K.,
  Ellis, R.\ S., Santiago, B.\ X., Surma, P.\ \& Griffiths, R.\ E.\ 
  1996a, \apjs, 107, 1
  
  \reference Abraham, R.\ G., Tanvir, N.\ R., Santiago, B.\ X., Ellis,
  R.\ S., Glazebrook, K.\ \& van den Bergh, S.\ 1996b, \mnras, 279, L47

  \reference Binney, J. 1992, ARAA, 30, 51
  
  \reference Binney, J., and Tremaine, S. 1987, Galactic Dynamics,
  Princeton University Press.
  
  \reference Blitz, L., Binney, J., Lo, K.~Y., Bally, J., and Ho,
  P.~T.~P. 1993, Nature, 361, 417
  
  \reference Bosma, A. 1991, in Warped disks and inclined rings around
  galaxies, ed. S. Casertano, P. Sackett and F. Briggs, 181
  
  \reference Brada, R.\ \& Milgrom, M.\ 1999a, \apjl, 512, L17
  
  \reference Brada, R.\ \& Milgrom, M.\ 1999b, \apj, 519, 590
  
  \reference Brada, R.\ \& Milgrom, M.\ 2000a, \apjl, 531, L21
  
  \reference Brada, R.\ \& Milgrom, M.\ 2000b, \apj, 541, 556
  
  \reference Burton W. B. 1998, in G. L. Verschuur, K. I. Kellermann
  (eds.), Galactic and Extragalactic Radio Astronomy, Springer-Verlag,
  Berlin, p. 295
  
  \reference Carney, B.\ W.\ \& Seitzer, P.\ 1993, \aj, 105, 2127

  \reference Chandrasekhar, S.  1969, Ellipsoidal Figures of
  Equilibrium, Yale University Press
  
  \reference Conselice, C.~J., Bershady, M.~A., and Jangren, A 2000,
  \apj, 529, 886
  
  \reference Debattista, V.~P. and Sellwood, J.~A.  1998, \apjl, 493,
  L5

  \reference Demleitner, M. 1998, \aap, 331, 485

  \reference Djorgovski, S.\ \& Sosin, C.\ 1989, \apjl, 341, L13

  \reference Drimmel, R. Smart, R.~L., and Lattanzi, M.~G. 1999,
  preprint (astro-ph/9912398)
  
  \reference Englmaier, P.\ \& Gerhard, O.\ 1999, \mnras, 304, 512

  \reference Hofner, P.\ \& Sparke, L.\ S.\ 1994, \apj, 428, 466

  \reference Henderson, A.~P., Jackson, P.~D. and Kerr, F.~J 1982,
  \apj, 263, 116

  \reference Garcia-Ruiz, I., Kuijken, K., Dubinski, J. 2000,
  preprint, (astro-ph/0002057)
  
  \reference Hernquist, L.\ 1990, \apj, 356, 359
  
  \reference Heyer, M.\ H., Brunt, C., Snell, R.\ L., Howe, J.\ E.,
  Schloerb, F.\ P.\ \& Carpenter, J.\ M.\ 1998, \apjs, 115, 241

  \reference Hunter, C., and Toomre, A. 1969, \apj, 155, 747
  
  \reference Ibata, R., Lewis, G.~F., Irwin, M., Totten, E., and
  Quinn, T. 2000, preprint (astro-ph/0004011)
  
  \reference Kuijken, K.\ 1998, ASP Conf.\ Ser.\ 136: Galactic Halos,
  349
  
  \reference Navarro, J.\ F., Frenk, C.\ S.\ \& White, S.\ D.\ M.\ 
  1997, \apj, 490, 493

  \reference Olling, R.~P. and Merrifield, M.~R., 2000, \mnras, 311,
  361
  
  \reference Pasha,I.~I.\ \& Polyachenko, V. L. \ 1993, Astr. Lett, 19, 1

  \reference Rudnick, G.\ \& Rix, H.\ 1998, \aj, 116, 1163
  
  \reference Rudnick, G. Rix, H.-W., and Kennicut, R. C. Jr. 2000,
  \apj, 538, 569
  
  \reference Scalo, J.~M. 1987, in Starbursts and Galaxy Evolution,
  ed. T. Thuan, T. Montmerle, and J. Van, Editions Frontieres, p.
  445-465
  
  \reference Sparke, L.\ S.\ \& Casertano, S.\ 1988, \mnras, 234, 873
  
  \reference Tremaine, S.\ 1993, AIP Conf.\ Proc.\ 278: Back to the
  Galaxy, 599

  \reference Weinberg, M.~D. 1994, \apj, 421, 481
  
  \reference Weinberg, M.\ D.\ 1998a, \mnras, 297, 101
  
  \reference Weinberg, M.\ D.\ 1998b, \mnras, 299, 499

  \reference Weinberg, M.~D. 2000a, preprint (astro-ph/0007275)
  
  \reference Weinberg, M.~D. 2000b, preprint (astro-ph/0007276)
  
  \reference de Vaucouleurs, G.\ \& Pence, W.\ D.\ 1978, \aj, 83, 1163

  \reference Vesperini, E.\ \& Weinberg, M.\ D.\ 2000a, \apj, 534, 598
  
  \reference Vesperini, E. and Weinberg, M.~D. 2000b, in preparation.
  
  \reference Zaritsky, D., Smith, R., Frenk, C., and White, S.~D.~M
  1997, \apj, 478, 39

\end{references}
\end{document}